\documentclass[11pt,eqsecnum,tightenlines,floats,aps,amsmath,
amssymb,nofootinbib,prd,noshowpacs,notitlepage,longbibliography]{revtex4}
\usepackage[utf8]{inputenc}
\usepackage{amsmath,amsthm,latexsym,amssymb,amsfonts,color,graphicx}
\usepackage{hyperref}

\def\be{\begin{equation}}
\def\ee{\end{equation}}
\def\ba{\begin{eqnarray}}
\def\ea{\end{eqnarray}}

\newcommand\nn{\nonumber}
\newcommand{\q}{\quad}

\begin{document}

\title{3D holography: from discretum to continuum}

\author{Valentin Bonzom${}^{1}$}
\email{bonzom@lipn.univ-paris13.fr}
\author{Bianca Dittrich${}^{2}$}
\email{bdittrich@perimeterinstitute.ca}

\affiliation{
  ${}^{1}$ LIPN, UMR CNRS 7030, Institut Galil\'ee, Universit\'e Paris 13, Sorbonne Paris Cit\'e,\\
  99, avenue Jean-Baptiste Cl\'ement, 93430 Villetaneuse, France\\
  ${}^{2}$Perimeter Institute for Theoretical Physics, 31 Caroline Street North
  Waterloo, Ontario Canada N2L 2Y5}

\begin{abstract}
We study the one--loop partition function of 3D gravity without cosmological constant on the solid torus with arbitrary metric fluctuations on the boundary. To this end we employ the discrete approach of (quantum) Regge calculus. In contrast with similar calculations performed directly in the continuum, we work with a boundary at finite distance from the torus axis.  We show that after taking the continuum limit on the boundary -- but still keeping finite distance from the torus axis -- the one--loop correction is the same as the one recently found in the continuum in Barnich et al. for an asymptotically flat boundary.  The discrete approach taken here allows to identify the boundary degrees of freedom which are responsible for the non--trivial structure of the one--loop correction.  We therefore calculate also the Hamilton--Jacobi function to quadratic order in the boundary fluctuations both in the discrete set--up and directly in the continuum theory. We identify a dual boundary field theory with a Liouville  type coupling to the boundary metric. The discrete set--up allows again to identify the dual field with degrees of freedom associated to radial bulk edges attached to the boundary. Integrating out this dual field reproduces the (boundary diffeomorphism invariant part of the) quadratic order of the Hamilton--Jacobi functional. The considerations here show that bulk boundary dualities might also emerge at finite boundaries and  moreover that discrete approaches are helpful in identifying such dualities.

\end{abstract}

\maketitle

\section{Introduction}

The AdS/CFT correspondence \cite{maldacena,witten,book} proposes a duality between gauge theories at the boundary of an asymptotically AdS spacetime and quantum gravity in the bulk. This correspondence has been turned into a powerful tool and has also been suggested to provide a definition of quantum gravity through the dual theory on the boundary.  

Other approaches such as loop quantum gravity \cite{lqg} and spin foams \cite{spinfoams} aim at defining directly a theory of quantum gravity without using an (asymptotic) background. An interesting question is therefore whether these non--perturbative approaches are compatible with AdS/CFT \cite{freidelAdS} or other forms of duality. For further work related to this question see  \cite{KIRILL,BonzomLivineIsing, Norbert}. This work explores this question in the context of 3D gravity without a cosmological constant. 

The exploration of AdS/CFT duality in the context of e.g. loop quantum gravity has been hindered by mainly two issues: Firstly the construction of models incorporating a negative cosmological constant is much more involved than models with vanishing or positive cosmological constants, but recent progress can be found in \cite{girelli,aldo}. These complications can be avoided by turning to a generalization of the AdS/CFT duality to asymptotically flat gravity, which has been considered in \cite{BarnichFlat}.  In particular \cite{Barnich1} computes the one--loop partition function with a torus boundary at asymptotic infinity, showing that it is given by the character of a representation of the BMS group \cite{BMS}, which is the asymptotic symmetry group of gravity without a cosmological constant. This shows also that the one--loop partition function gives actually the exact result, i.e. that all higher order corrections vanish (for asymptotically flat boundary data).

Secondly and more importantly, it is very involved and to some extend goes against the philosophy of background independent approaches, to impose asymptotic boundary conditions. Indeed, partition functions for \emph{finite} regions are central in formulating and understanding the theory \cite{oeckl, RovelliHJ, bd12, bd14} and do encode the complete dynamics of the theory \cite{PerezRovelli}.  Such partition functions for finite regions are actually unavoidable in non--perturbative approaches: the size of the region is encoded in the boundary data on which the partition function depends.  Thus an asymptotic limit means to consider the partition function only for certain, very special types of boundary data, which do not encode the full dynamics of the theory.  

The aim we pursue here is to obtain the partition function for boundary data describing a finite region. In the example described here, 3D flat gravity with a torus boundary, we will nevertheless find a result consistent with the one expected due to the BMS symmetry of flat gravity and we will also identify a dual field theory, generalizing results  found for AdS gravity   \cite{skenderis,carlip} to the flat case. 

As mentioned one aim pursued here is to prepare the ground for calculations of the (finite region) partition functions in non--perturbative approaches. These often involve discretizations and we thus turn to a discrete approach, Regge calculus \cite{regge}. Here it is so far not possible to treat the quantum partition function analytically for the full non--linear theory -- at least not in the incarnation known as quantum Regge calculus \cite{qregge} -- however the linear theory can be straightforwardly handled and -- at least for the asymptotic case -- is expected to give the exact result.  3D gravity and also 3D (classical) Regge gravity is a topological theory -- there are no propagating degrees of freedom.   To make it also a topological, and for 3D Regge calculus, triangulation invariant theory in the quantum realm, one needs to identify the correct path integral measure. Such a measure is not known for the original quantum Regge calculus, but has been identified for the linearized theory in \cite{Meas}.  

The Ponzano--Regge model \cite{PR} can be seen as a quantization of Regge calculus based on a first order formulation of gravity (which happens to coincide with $SU(2)$ BF theory). As it is (at least formally) invariant under changes of triangulation \cite{TorsionPR}, it also has a triangulation invariant measure built in. In the semiclassical limit it moreover coincides with the one found in \cite{Meas}.  The Ponzano--Regge model is the first example of a spin foam model. We expect the  work in this paper to help consider the partition function for the Ponzano--Regge model on the solid torus and to examine possible dual field theories in a fully non--perturbative set--up. 

We can thus make use of the fact that 3D Regge calculus is a topological field theory. This allows us to work with the coarsest bulk triangulation possible, which simplifies the calculations drastically and moreover will lead to a very geometric interpretation of the result.  Another issue is the boundary triangulation. On the one hand this can be seen as a regulator. Indeed we will see that, despite working with a topological theory, the Hamilton--Jacobi functional can be described by a (dual) field theory with propagating degrees of freedom on the boundary.  To this end we have to take the continuum limit for the boundary discretization. On the other hand the discrete boundary can also be seen as a particular choice of boundary data for the continuum theory, where the boundary metric is piecewise flat \cite{bd12}. In this case the dual field theory comes in the form of a lattice theory, does however give the exact result for this kind of piecewise flat boundary data.

This work will show that discrete approaches are indeed useful for the exploration of holography.  The calculation of the one--loop determinant performed here is much simpler than in the continuum and makes direct use of the topological nature of the theory.  As mentioned a crucial issue is to be able to work at finite boundary, for which we here provide a proof of concept. Often one rather employs an asymptotic expansion, e.g. the Fefferman--Graham expansion \cite{FG}, starting from the metric at infinity, for instance in \cite{skenderis,carlip}. In contrast, here we identify the induced boundary theory for a finite boundary (for linearized gravity). Indeed in the case of 3D flat gravity it will turn out that it is not even necessary to take the asymptotic limit to validate the one--loop correction.   

Moreover we provide an explicit example of a discrete model leading to a boundary field theory. A related concept are MERA tensor networks, designed to describe conformal field theories \cite{MERA}. Recently a connection of these networks to AdS/CFT duality (or other forms of duality) has been  proposed and explored \cite{AdSMERA}. The discretization employed here can also be seen as a generalized tensor network (in particular as a decorated tensor network introduced in \cite{dectnw}). The dynamics incorporates directly 3D gravity and leads indeed to a field theory on the boundary. Due to the topological nature of the theory, the connectivity of the network, given by the choice of (bulk) discretization, is not important. This feature does reflect the diffeomorphism invariance of the theory \cite{improved,perfect}, an issue that so far is left open in the context of the MERA approach to AdS/CFT. As we will discuss we can choose a discretization that emulates a MERA network and thus provide a model with a direct connection to gravity.

~\\
In Section \ref{reviewB} we provide a short review of the work \cite{Barnich1}, which computes the one--loop partition function for 3D gravity with vanishing cosmological constant and torus boundary. To prepare for the corresponding computation in a discrete set--up and for finite boundary we summarize all the necessary ingredients of Regge calculus in Section \ref{Sec:Regge}. A particular important role will be played by the diffeomorphism invariance of the theory, discussed in Section \ref{Sec:Diff}.  We construct the partition function for Regge calculus in Section \ref{Sec:PD}, which we then calculate in Section \ref{Sec:Int}. In particular Section \ref{Sec:Loop} discusses the one--loop determinant and gives a geometric interpretation of its structure. In Section \ref{Sec:Eff} we extract the Hamilton--Jacobi functional, or effective boundary action (for the linearized theory and) for inhomogeneous boundary data and we perform the continuum limit in Section \ref{Sec:Climit}.  In Section \ref{Sec:L} we conjecture and test a dual field theory on the boundary. This dual field theory is coupled to the boundary curvature and integrating out the dual field will give the part of the effective boundary action which is invariant under linearized boundary diffeomorphisms. This part is also responsible for the dependence of the one--loop determinant on the topological parameters of the torus. Finally we confirm our calculation of the effective boundary action in the discrete by computing the effective boundary action directly in the continuum theory in section \ref{Sec:Cont}.  In section \ref{outlook} we discuss two issues left for further work: one is the use of different discretizations, in particular one that relates to MERA tensor networks. Such discretizations also allow to perform a continuum limit in the bulk which would be necessary to consider for non--topological theories, for instance if one considers gravity coupled to matter. Another issue discussed is the use of alternative boundary terms and conditions. We close the paper with a discussion in section \ref{discussion}.

\section{The partition function in the continuum}\label{reviewB}

In \cite{Barnich1}  the one--loop partition function of three--dimensional gravity without cosmological constant is computed.  The boundary has the topology of the torus and the boundary data coincide with asymptotic boundary condition prescribing a geometry known as thermal spinning flat space \cite{ThermalSpinning}.  This is a  3D flat spacetime
\ba\label{metric0}
ds^2&=& dr^2  + dt^2+ r^2 d\phi
\ea
with periodic identification $(r,t,\phi)\sim (r,t+\beta,\phi+\gamma)$ in addition to the usual identification $\phi\sim \phi+2\pi$ for the angular variable. If we consider the spacetime for $0\leq r\leq R$ we obtain a solid torus with contractible cycles given by $t=\text{const},\, r=\text{const}$ and non--contractible cycles along the  $\phi=\text{const}., \, r=\text{const}$ lines. The torus can be obtained by identifying the top and bottom discs of a cylinder of height $\beta$, with a twisting angle (or angular potential) $\gamma$.   

To compute the one--loop partition function one needs to evaluate the (Euclidean) Einstein--Hilbert action with boundary term
\ba\label{EHaction}
S_{EH}&=&-\frac{1}{16\pi G} \int d^3 x \sqrt{g} R \,\,+\,\, B
\ea
on the solution. In this work we choose as boundary term the standard Gibbons--Hawking--York term \cite{GHY}
\ba \label{GHY}
B_{GHY}&=& -\frac{1}{8\pi G} \int d^2 y \sqrt{h} K 
\ea 
with $h$ the induced metric on the boundary and $K=h^{\mu\nu}K_{\mu\nu}$ the trace of the intrinsic curvature. Note that \cite{Barnich1} uses a boundary term $B'=\frac{1}{2}B_{GHY}$ instead. The use of this boundary term has been justified in \cite{detournay} in the asymptotic limit $R\rightarrow \infty$ and comes from the fall-off conditions on the metric coefficients at infinity. It is this choice $B'$ of boundary term that realizes the relation of the partition function to the BMS vacuum character. We will however stick here to $B_{GHY}$. The reason is our strategy to work with a boundary at finite radius, for which $B_{GHY}$ is the correct boundary term. In addition, Regge calculus deeply relies on using the $B_{GHY}$ term: the actions for single building blocks coincide with the continuum action evaluated on chunks of flat spacetime with piecewise flat boundaries. We will also see that the one--loop correction is not affected by the choice of boundary term.
We will comment on the possible use of $B'$ in the context of Regge calculus in the outlook section \ref{outlook}.

The bulk action vanishes on solutions and thus the zeroth order of the partition function is given by the boundary term. The trace of the extrinsic curvature for a $r=R$ surface in the  metric \eqref{metric0} is given by $K=1/R$, which together with $\sqrt{h}=R$ leads to a term proportional to the area of the torus at unit radius $r=1$, and hence a $R$-independent, finite result. Together with all constants we have
\ba
B_{GHY}\,=\, -\frac{\beta}{4G} \q .
\ea
 
The one--loop correction  is given by evaluating the determinant of the second order fluctuation matrix of the action, expanded around the given background.  Such a one--loop correction has been found first for the AdS case \cite{giombi} using heat kernel techniques. In \cite{Barnich1} the same strategy is followed: after including the appropriate gauge--fixing and ghost terms, the one--loop partition correction can be written as
 \ba
 S^{(1)} \,=\, -\frac{1}{2} \ln \det \Delta_{(2)} + \ln \det \Delta_{(1)} - \frac{1}{2} \ln \det \Delta_{(0)}
 \ea
where $\Delta_{(i)}$ for $i=2,1,0$ are the Laplacians for the traceless transverse, the vector, and the scalar mode of a symmetric $3\times 3$ matrix.

The determinants can be found using the heat kernel approach. To this end one needs to find the solutions of the heat kernel equations corresponding to the Laplacians $\Delta_{(i)}$ on the background under considerations. The heat kernels are known for flat space of topology $\mathbb{R}^3$. One can obtain the heat kernels for the thermal spinning spacetime by applying the method of images, which sums the $\mathbb{R}^3$ kernels over the images of the discrete symmetry group generated by the mapping $(r,t,\phi)\rightarrow (r,t+\beta,\phi+\gamma)$. Furthermore one performs a regularization by adding a mass term to the heat kernels, which is then removed in the final result.

Collecting the  three contributions for the various modes, this sum (originating from the sum over images) takes the form
\ba \label{ContinuumS1}
S^{(1)}=\sum_{n=1}^\infty \frac{1}{n}\left( \frac{q^{2n}}{1-q^n} +  \frac{\bar{q}^{2n}}{1-\bar{q}^n} \right)
\ea
where $q=e^{i(\gamma+i\varepsilon)}$ with a further regulating parameter $\varepsilon$.  After rewriting the exponential of this sum, the one--loop partition function can  be expressed as
\ba\label{Z1}
Z(\beta,\gamma) \,=\, e^{-\frac{B}{\hbar}} \prod^\infty_{k=2} \frac{1}{|1-q^k|^2} \, + {\cal O}(\hbar) \q .
\ea
For the choice of boundary term $B=\tfrac{1}{2}B_{GHY}=-\frac{\beta}{8G}$ this reproduces the 3D vacuum BSM character constructed in \cite{oblak}. 

Our aim is to reproduce the one--loop correction in the setup of Regge calculus, using a boundary at finite radius. Whereas the nature of 3D gravity as a topological theory is rather hidden in the continuum computation, we will make direct use of it. This will make the structure of the result  (\ref{Z1}) transparent. It shows that the one--loop contribution can be understood as arising from degrees of freedom residing at the boundary of the spacetime, which are equipped with a kinetic term $\partial_t^2$. The one--loop correction arises as the determinant of $\partial^2_t$ on the boundary torus, where the integer $k$ in (\ref{Z1}) is found to be the momentum (Fourier) mode in angular direction. We will also show that the product in (\ref{Z1}) starts with $k=2$ due to diffeomorphism symmetry. It indeed cancels out the $k=0$ and $k=\pm 1$ modes. These findings motivate us to search in Section \ref{Sec:L} for a dual field theory describing the boundary fluctuations (after integrating out the bulk variables), which can be understood to be responsible for the one--loop correction.

\section{Review Regge calculus}\label{Sec:Regge}

In this section we provide a short review of Regge calculus and collect all material necessary to be able to calculate the partition function for the linearized theory. An important role is played by a discrete notion of diffeomorphism symmetry extant  in 3D Regge calculus, which we also present.

\subsection{The Regge action and equations of motion}\label{sec:ReggeAction}

In Regge calculus one replaces the smooth manifold ${\cal M}$ with a triangulation ${\cal T}$. The metric data are then encoded by an assignment of length variables $l_e$ to the edges $e$ of the triangulation ${\cal T}$. This specifies a piecewise flat and linear geometry for ${\cal T}$. The solutions of the theory are determined by varying the 3D Regge action
\ba\label{regge1}
S_R&=&
- \frac{1}{8\pi G}\sum_{e \subset {\cal T}^\circ}  l_e \epsilon_e  \,\,-\,\,  \frac{1}{8\pi G}\sum_{e \subset \partial{\cal T}} l_e \omega_e \q .
\ea
Here the first sum is over all bulk edges of ${\cal T}$ and involves the deficit angles
\ba\label{regge2}
\epsilon_e(l_{e'})&=& 2\pi- \sum_{\sigma  \supset e} \theta_e^\sigma (l_{e'}) \q ,
\ea
which measure the curvature\footnote{The deficit angle agrees with the angle of rotation that a vector receives after being parallel-transported fully around the edge. The parallel-transport can be defined locally on each pair of glued tetrahedra, as such a pair can be embedded into flat spacetime.} concentrated on a given edge $e$. The  angle $\theta_e^\sigma$ is the interior dihedral angle at the edge $e$ in the tetrahedron $\sigma$. It is a function of the six edge lengths of $\sigma$.

The second sum in (\ref{regge1}) corresponds to the Gibbons--Hawking--York  boundary term in (\ref{EHaction}) and involves the boundary angles
\ba\label{regge3}
\omega_e(l_{e'})&=& \pi- \sum_{\sigma  \supset e} \theta_e^\sigma (l_{e'}) \q ,
\ea
which now measure the extrinsic curvature. This (discrete equivalent of the Gibbons--Hawking--York) boundary term makes the variational principle well defined for the action \eqref{regge1} with fixed boundary edge lengths \cite{sorkinBdry}.

Indeed, due to the Schl\"afli identity 
\ba\label{schlaefli}
\sum_{e \in \sigma}  l_e  \delta \theta_e^\sigma \,=\,0
\ea
one can show that the variations $\delta \epsilon_e$ vanish, if varying the Regge action (\ref{regge1}). One thus obtains the equations of motion
\ba\label{regge5}
0 &\stackrel{!}{=}&8\pi G\, \delta_e S_R \,=\,- \epsilon_e \q, 
\ea
demanding -- as one expects in 3D gravity -- vanishing curvature. As we will later comment these equations might not fix the edge length in the bulk uniquely, a feature that leads to a discrete notion of diffeomorphism symmetry \cite{rocek, dittrich08}. 

This non--uniqueness does however not matter for finding Hamilton--Jacobi functional (or Hamilton's principal function or effective boundary action), that is for evaluating the action on the solution. The equation of motions make the bulk term vanish and we are left with the evaluation of the boundary term
\ba\label{regge6}
{S_R}_{|\text{sol}} = - \frac{1}{8\pi G}\sum_{e \subset \partial{\cal T}} l_e \, \omega_e({l_{e'}}_{|\text{sol}}  )   \q .
\ea
Thus Hamilton's principal function coincides with the (integrated) extrinsic curvature of a piecewise flat  2D surface embedded into flat 3D space. Note that in this case Hamilton's principal function is independent of any choice of bulk triangulation: the bulk triangulation is only needed as an auxiliary construction to determine the embedding of the 2D boundary surface into flat 3D space. Triangulation independence of local actions leads to theories without  propagating (non--topological) degrees of freedom, that is correspond to perfect discretizations of topological theories (see eg. \cite{bd12review,perfect}).  Thus Regge calculus is a perfect discretization of 3D gravity \cite{improved,he}, that is it captures exactly the continuum properties of this (topological) theory.

Let us also remark that the embedding of the 2D surface into flat spacetime and therefore Hamilton's principal function might not be uniquely defined. Here we refer to relative embedding information: that is even if we fix the position of all the neighbouring vertices  of a certain vertex $v$, the position of $v$ itself might not be unique. These are however discrete ambiguities, and are indeed connected to the sum over spacetime orientations in the Ponzano--Regge action.\footnote{As an example one can consider a boundary triangle that is further subdivided into three triangles. Given the six edge lengths involved there will be (in general) two possibilities to realized the embedding. One case corresponds to gluing a tetrahedron onto the outer side of the triangle and therefore introducing a bulge into the exterior. The other case is to glue the same tetrahedron onto the inner (bulk) side of the boundary triangle. See \cite{hoehn2} for a detailed discussion of how to understand such gluings as (time) evolution of the boundary. In the continuum, these choices correspond to positive and negative lapse in the evolution of the boundary. They are therefore also related to the question of summing over orientations for the partition function \cite{antispacetime}. The work \cite{freidelAdS} includes a discussion of this feature in the context of AdS/CFT duality.}

Note that we can rewrite the Regge action as a sum over tetrahedra (plus  terms that will not contribute to the Hessian of the action)
\ba\label{regge7}
8\pi G\, S_R &=& -\sum_{e \subset {\cal T}^\circ} 2\pi  l_e\,\,-\,\, \sum_{e \subset \partial{\cal T}} \pi l_e   \,\,+\,\, \sum_{\sigma \subset {\cal T}} S_\sigma
\ea
where
\ba\label{regge8}
S_\sigma  - \sum_{e\in \sigma} \pi l_e &=& \sum_{e\in \sigma} l_e \theta_{l_e} - \sum_{e\in \sigma} \pi l_e
\ea
gives Hamilton's principal function for the boundary data of a tetrahedron. This  Hamilton's principal function does coincide with the continuum Hamilton's function if one understands the boundary metric to be piecewise flat and extrinsic curvature to be distributional concentrated on the edges of the tetrahedron.  Thus Hamilton's principal function for a tetrahedron is also invariant under any (bulk) refinements of the tetrahedron. 

The linearized theory is described by expanding the Regge action to quadratic order in perturbations $l_e=L_e + \lambda_e$:
\ba\label{regge9}
S^{(2)}&=& \frac{1}{2}   \sum_{e,e'} {\frac{\partial S_R}{\partial l_e \partial l_{e'}} }_{| l_e=L_e}  \lambda_e \,\lambda_{e'}  \nn\\\
&=&
 \frac{1}{16\pi G} \sum_{e,e'} \,\,     \sum_{\sigma  \supset e,e'}\!\! \frac{\partial \theta_e^\sigma}{\partial l_{e'} }_{|l_e=L_e} \,  \lambda_e \lambda_{e'}  \q .
\ea
Here  Schl\"afli identity (\ref{schlaefli}) guarantees that $\partial \theta_e^\sigma / \partial l_{e'} = \partial \theta_{e'}^\sigma / \partial l_e$. 

Bulk triangulation independence of the full action (or rather Hamilton's principal function) does also extend to the linearized case \cite{he,Meas}.

\subsection{Diffeomorphism and scaling symmetry}\label{Sec:Diff}

The topological nature of 3D continuum gravity and also its discrete incarnation in the form of Regge calculus is due to gauge symmetries which reduce the field variables to a few topological degrees of freedom. Gauge symmetries lead to non--uniqueness of solutions. In the case of Regge calculus we can easily understand this non--uniqueness: given a solution with bulk vertices we know that this solution is 3D flat. We obtain a physically equivalent solution by displacing the position of a given bulk vertex in the embedding flat 3D space. This displacement will lead to a change in the length of the edges connected to the vertex, but clearly not to a change of the geometric properties of the bulk space. 

These symmetries have been discussed in  \cite{rocek, hamber-williams,louapre, dittrich08}. For 4D the status of gauge symmetries depends on the solution: a vertex embedded in a flat piece of space can be displaced whereas it has been shown in \cite{dittrich08} that diffeomorphism symmetry is broken for solutions with curvature.

The diffeomorphism symmetry in the discrete is closely connected to (bulk) triangulation independence \cite{perfect,bd12review}: If we can displace vertices one can also move a vertex on top of another vertex which leads to a coarse--graining of the triangulation. This shows that the coarse--graining of the triangulation does not change Hamilton's principal function. 

For the linearized action the gauge symmetries result in null vectors for the associated Hessian matrix. We thus expect three null vectors for every bulk vertex. 

There is a further (global) null vector we can expect, resulting from the Schl\"afli identity.   As one can see from (\ref{regge9}) each tetrahedron $\sigma$ contributes  
\ba\label{diffeo1}
H_{ee'}^\sigma:= \frac{\partial \theta_e^\sigma}{\partial l_{e'} }   \q , \q\q H_{ee'}=\sum_{\sigma \supset e,e' } H_{ee'}^{\sigma}  \q .
\ea
Due to the Schl\"afli identity (\ref{schlaefli}) the Hessian for a tetrahedron has a null vector which extends to a null vector for the full triangulation Hessian (including boundary terms!):
\ba
\sum_{e' \subset \sigma} H_{ee'}^\sigma l_{e'} \,=\, 0  \q \q  \Rightarrow  \q\q \sum_{e'} H_{ee'} l_{e'} \,=\,0  \q .
\ea 
The null vectors correspond to a scaling symmetry $\lambda_e \sim L_e$. Indeed, going back to the full theory, if one multiplies all edge lengths of the solutions with a scaling factor, one again obtains a solution (with also rescaled boundary data). The full Hamilton--Jacobi functional and also action is however not invariant -- it rather is multiplied by the same scaling factor. One can argue \cite{dfs} that the combination of null vectors corresponding to scaling symmetry  for each tetrahedron lead to the null vectors corresponding to diffeomorphism symmetry.

\subsection{Quantum Regge calculus: The path integral measure}\label{Sec:Meas}

In quantum Regge calculus  \cite{qregge} one considers the path integral
\ba \label{FormalPFRegge}
Z\,=\, \int {\cal D} \mu (l) \exp\left(-\frac{1}{\hbar} S_{R} \right)
\ea
integrating over the lengths of the bulk edges  in the triangulation with some measure $\mu(l)$.  There has been considerable debate in the literature on the measure for quantum Regge calculus \cite{Reggemeasure}.  One difficulty that makes any analytical calculations hard are the triangle inequalities that make determining the integration range very complicated. We avoid this issue by considering linearized Regge calculus and integrating the length perturbations over $\lambda \in \mathbb{R}$ as in standard perturbative quantum field theory.

In 3D, given the invariance properties of the classical action, one can demand that the same invariances should hold at least to one--loop order (for the linearized theory). Indeed it has been shown in \cite{Meas} that invariance under changes of the bulk triangulation does  fix the measure of 3D Regge calculus uniquely. This measure term does furthermore coincide with the asymptotics \cite{asympPR} of the Ponzano--Regge model \cite{PR}, which can be understood as a quantized (first order) version of Regge calculus. This measure is given by
\ba\label{Rmeasure}
{\cal D} \mu (l) \,=\,  \prod_{\sigma} \frac{1}{\sqrt{12 \pi V_\sigma}} \prod_{ e \in \rm{bulk}} \frac{ l_e\, d l_e}{\sqrt{8 \pi G \hbar}}  \prod_{ e \in \rm{bdry}} \sqrt{ \frac{ l_e}{\sqrt{8 \pi G \hbar}}  } \q .
\ea
Here invariance of the measure means, that the form of the measure does not change if one integrates out edge variables. This integration process can be organized locally and interpreted as changing and in fact coarse--graining the triangulation. There are two such local changes, the so--called $4-1$ and $3-2$ Pachner moves, which generate (together with their inverses) all possible (bulk) changes of the triangulation. For the $3-2$ move one integrates out one edge, and one can show explicitly that the form of the measure (\ref{Rmeasure}) does not change. For the $4-1$ Pachner move one integrates out four edges. The corresponding Hessian matrix has one negative eigenvalue (which thus would lead to a divergence and reflects the conformal mode problem of gravity) and three null eigenvalues, corresponding to the vertex--displacement gauge symmetry, that one expects. The negative eigenvalue can be dealt with, as in the continuum \cite{GibbonsHawking}, by `analytically continuing' this mode to have a positive eigenvalue. To deal with the three null eigenvalues one proceeds by (a) integrating out one edge (which does not coincide with a null vector) and (b) removing a measure factor that corresponds to the measure over the gauge orbit \cite{aristide3D}. This latter measure factor is only determined up to numerical constants. To achieve form invariance of (\ref{Rmeasure}) one has to choose to remove
\ba
\frac{1}{(8\pi G \hbar)^{3/2}}\frac{1}{2\pi}  \prod_{a=1,2,3} d x_p^a \,=\, \frac{1} {12\pi V_{\bar{1}}} \prod_{i=2,3,4} \frac{l_{0i}}{\sqrt{8\pi G\hbar}} d\lambda _{0i}
\ea
where $\prod_p dx^a_v$ is the Euclidean measure over the position $p$ of a vertex in $\mathbb{R}^3$, which gives the gauge orbit corresponding to the vertex displacement symmetry. Here one uses  the identity \cite{aristide3D, Meas}
\ba
\prod_{a=1}^d dx^a_0\,=\, \frac{1}{d! V_{\sigma} } \prod_{i=1}^d l_{0i} dl_{0i}
\ea
between the $d$--dimensional Euclidean measure associated to the cartesian coordinates of the vertex $v=0$ and the edge lengths $l_{0i}$ of a $d$--simplex with vertices $v=0,1,\ldots,d$ and volume $V_{\sigma}$.

The important point is, that the existence of this triangulation  invariant measure in 3D allows us to choose any bulk triangulation for the evaluation of the partition function to one--loop order, and we will choose the coarsest one possible.  Such a (local) measure does exist only in 3D. For 4D Regge calculus one can show that no local measures exist that reproduce the amount of triangulation invariance extant in the classical action, which is invariant under $5-1$ and $4-2$ Pachner moves, but not under $3-3$ Pachner moves \cite{Meas}.

\section{Partition function in the discrete set--up}\label{Sec:PD}

Having collected all the necessary preliminaries, we are going to define and compute the partition function for linearized Regge calculus. To this end we have to (a) choose a discretization and (b) find the linearized Regge action for this triangulation and the background geometry at hand. The linearized Regge action needs to be brought into a convenient form -- here by a `twisted' lattice Fourier transform, so that the bulk variables can be easily integrated out. Linearized Regge calculus on a regular lattice has been first considered in \cite{rocek} (see also \cite{dfs} for a slightly more general set--up). Although it did not consider a boundary, the main steps we use have been developed in \cite{rocek}.

\subsection{ The triangulation}

Given the bulk triangulation independence of 3D Regge calculus it is sufficient to choose for our calculation a very coarse bulk triangulation. However we might have to take the continuum limit for the boundary triangulation and should therefore allow for a sufficiently general triangulation to achieve it.

\begin{figure}
\includegraphics[scale=0.7]{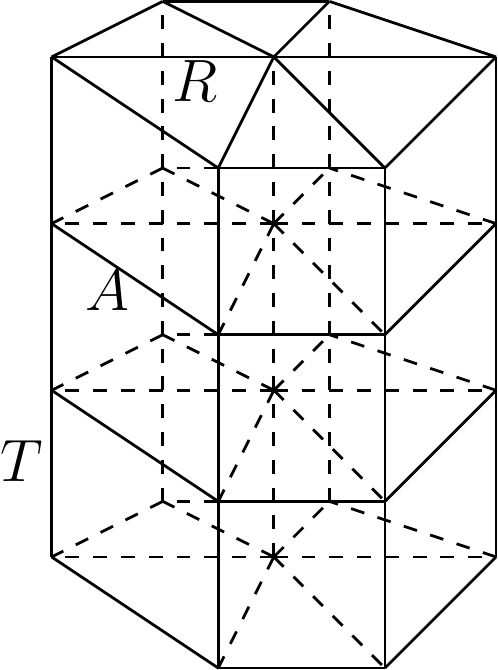} \q\q\q\q\q\q\q\q
\includegraphics[scale=1]{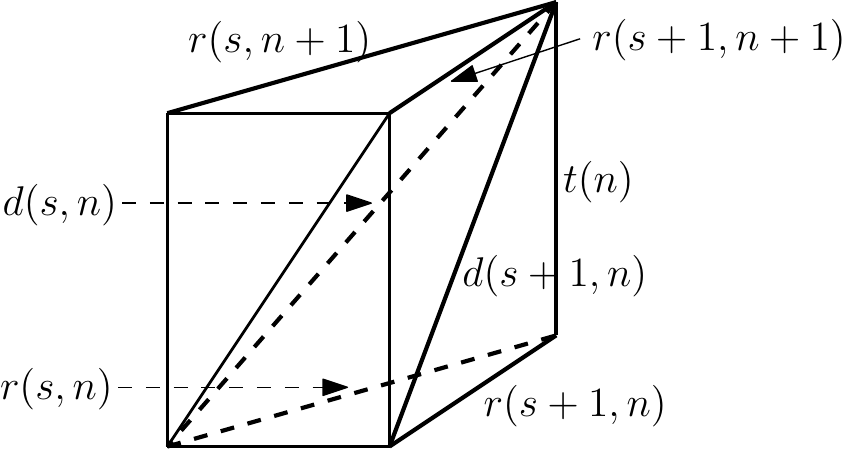}
\caption{\small The torus is cut open and the resulting cylinder is cut into prisms. Each prism is triangulated with three tetrahedra.  On the right panel, the bulk edges are named and labeled by the time step $n$ and angular position $s$.
\label{fig_1}}


\end{figure}

The spacetime we are considering has the topology of a solid torus. We cut this torus to a cylinder of height $\beta$ and slice it into $N_T$ cylinders of smaller height $T$. Note that in gluing back the cylinder to a solid torus we have to take into account the angular potential/twist parameter. Each cylinder is furthermore cut along $N_A$ radial lines of length $R$, and approximated by $N_A$ prisms (their boundary becomes piecewise linear), see Figure \ref{fig_1}.  Each of those prisms can be triangulated with three tetrahedra, see Figure \ref{fig_1}. 

\begin{figure}[htb]
\includegraphics[scale=0.7]{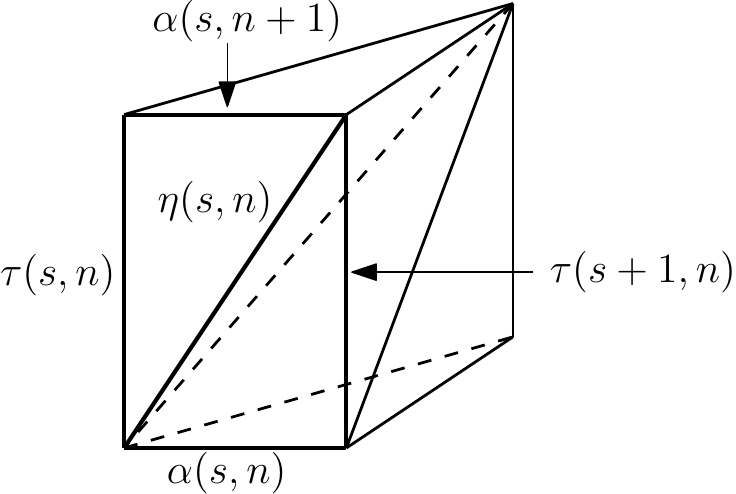}
\caption{\small  The boundary variables on a prism.}\label{fig_2}
\end{figure}

This introduces 6 types of edge lengths. There are three types of bulk edges:
\begin{itemize}\parskip-2mm
\item The bulk radial edges, forming four edges of the prism, whose background length we denote by $R$ and whose small  deviation we denote by $r$.  That is we write the radial length variables as $l_R=R+r$.
\item The bulk edges subdividing two sides of the prism diagonally with background length $\sqrt{R^2+T^2}$. We will denote the corresponding  fluctuation variable by $d$.
\item The edges that make up the axis of the cylinder (i.e. of radial coordinate $r=0$) of background length $T$ and associated fluctuation variable $t$.
\end{itemize}
We have furthermore three types of boundary edges, see Figure \ref{fig_2}.
\begin{itemize}\parskip-2mm
\item The edges in angular direction, whose background length and  fluctuations are denoted by $A$ and $\alpha$ respectively.
\item The edges in time direction with background length $T$  and  fluctuation variables $\tau$.
\item The edges in diagonal direction with background length $\sqrt{A^2+T^2}$ and fluctuation variable $\eta$.
\end{itemize}

This choice of background length variables make the deficit angles at almost all inner edges vanish. The exception are the edges making up the time axis in the centre of the cylinder. Requiring a vanishing deficit angle relates $A,R$ and $N_A$, the number of prisms making up one time slice by
\ba\label{t1}
x&:=&\frac{A^2}{2R^2} \,\,\stackrel{!}{=}\,\, 1-\cos{\frac{2\pi}{N_A}}  \q .
\ea

The boundary of the solid torus is triangulated as a regular (rectangular) lattice. We give coordinates $(s,n) \in [0,\dotsc,N_A-1]\times [0,1,\ldots N_T-1]$ to the vertices of this lattice. On the cylinder (oriented in time direction) we also orient all edges so that they point either to the right, up or right--up. We orient the bulk radial and diagonal edges to point to the centre of the cylinder and the time axis edges to point up. We then attach the source vertex coordinates to the  fluctuation variables, e.g. we write $\tau(s,n)$ or $r(s,n)$. To the time axis variables we just attach the time step variable $n$, i.e. we use $t(n)$.

We can now make \eqref{FormalPFRegge} more precise. The quantity of interest in this study is
\begin{equation} \label{PF}
Z(\{\tau(s, n), \alpha(s, n), \eta(s, n)\}) = 
\int \prod_{n} dt(n) \prod_{s, n} dd(s, n) dr(s, n)\ \mu(l)\ \exp\left( -\frac1{\hbar} S_R(l)\right)
\end{equation}
as a function of the boundary fluctuations $\{\tau(s, n), \alpha(s, n), \eta(s, n)\}$. Here $\mu(l)$ is the density of the measure $\mathcal{D}\mu(l)$, \eqref{Rmeasure} evaluated on the background triangulation,
\begin{equation}
\mu(l) = \frac{R^{N_A N_T} T^{N_T} (R^2 + T^2)^{N_A N_T}}{(8\pi G \hbar)^{N_T(2 N_A + 1)/2}}\,\sqrt{\frac{A^{N_A N_T} T^{N_A N_T} (A^2 + T^2)^{N_A N_T/2}}{(8\pi G \hbar)^{3N_A N_T/2}}}\,\frac1{(12\pi V_\sigma)^{3N_A N_T/2}}\q ,
\end{equation}
with
\begin{equation} \label{Volume}
V_{\sigma}= \frac{1}{12} ATR \sqrt{4-2 x}
\end{equation}
begin the (background) volume of each tetrahedron (it is the same for all three types of tetrahedra shown in Figure \ref{fig_1}). We will evaluate \eqref{PF} through an expansion of the Regge action to second order in the bulk and boundary fluctuations. The calculation gives a result of the form
\begin{equation} \label{PF2}
Z(\{\tau(s, n), \alpha(s, n), \eta(s, n)\}) = e^{-\frac1{\hbar} S_{R|\text{sol}}}\ D\ e^{- F(\{\tau(s, n), \alpha(s, n), \eta(s, n)\})}\q,
\end{equation}
where $S_{R|\text{sol}}$ is the action evaluated on the background solution, $D$ the 1--loop determinant and $F$ the part of the Hamilton--Jacobi functional which is linear and quadratic in the boundary fluctuations. We will evaluate $F$ by integrating the bulk fluctuations which are coupled to the boundary fluctuations.

\subsection{The Hamilton--Jacobi functional to zeroth and first orders in the boundary fluctuations}

We start with evaluating the Regge action (\ref{regge1}) on the background solution. As explained in (\ref{regge6}) we need to evaluate the boundary term
\ba
{S_R}_{|\text{sol}}  \,=\, - \frac{1}{8\pi G}\sum_{e \subset \partial{\cal T}} l_e \, \omega_e({l_{e'}}_{|\text{sol}}  )  \q .
\ea
For our background triangulation the only non--vanishing boundary angles reside at the boundary edges in time direction, for which $\omega_e=\frac{2\pi}{N_A}$. There are $N_A \times N_T$ such edges, which have length $T$. To match the torus boundary geometry with the continuum we need $N_T \times T =\beta$. We therefore obtain
\ba\label{S0R}
{S_R}_{|\text{sol}} \,=\, - \frac{\beta}{4G}
\ea
as in the continuum. To obtain this result we do not need any kind of continuum limit. This will change for the one--loop correction. We can also write $S_{HJ}^{(0)} = - \frac{\beta}{4G}$ where $S_{HJ}^{(0)}$ is the Hamilton--Jacobi functional at zeroth order in the boundary fluctuations.

We look at the variation of the Regge action \eqref{regge1} as in section \ref{sec:ReggeAction}, but this time we introduce variations of the boundary lengths. Denoting $\delta l_e$ the variation of the edge length $l_e$,
\begin{equation}
\delta S_R = -\frac1{8\pi G} \Bigl(\sum_{e\subset \mathcal{T}^\circ} \delta l_e\,\epsilon_e + \sum_{e\subset \partial \mathcal{T}} \delta l_e\,\omega_e\Bigr).
\end{equation}
Evaluating on the flat solutions, $\epsilon_e = 0$ for all interior edges, makes the bulk term vanish. Moreover, since the only non-vanishing boundary angles are those hinged along the time direction, which we denote $\omega(s,n)$,
\begin{equation}
8\pi G \,\delta S_{R|\text{sol}} = -\sum_{s,n} \tau(s,n) \omega(s,n) = - \frac{2\pi}{N_A} \sum_{s, n} \tau(s, n),
\end{equation}
where the last equality comes from $\omega(s, n) = \frac{2\pi}{N_A}$ for our choice of discretization and background. Here $\delta S_{R|\text{sol}} = S_{HJ}^{(1)}$ is the term of the Hamilton--Jacobi functional linear in the fluctuations.

\subsection{The  Hessians and their null vectors}

\subsubsection{Hessian for one prism}
As explained in section \ref{Sec:Regge} the Regge action and therefore its Hessian $H_{ee'}$ is additive over tetrahedra. Indeed we can consider the Hessian for any building block, e.g. for the prisms, which are our basic (uniform) building blocks.  The computation of this Hessian is explained in further detail in the Appendix \ref{sec:appendixA}. For the Hessian of a prism we obtain a matrix of the form\footnote{A similar scaling behaviour for the Regge action based on a regular cubical lattice has been observed in \cite{dfs}. In this case the matrix $M$ would not depend on any length parameters.}
\ba
H^{\rm pr}_{ee'} &=& \sum_{\sigma \in {\rm pr}} \sum_{\sigma \supset e,e'}  \frac{\partial  \theta^\sigma_{e'} }{\partial l_e} \,\,=\,\,\frac{L_e L_{e'}}{6 V_\sigma}  M^{\rm pr}_{ee'} (x)
\ea
where $M_{ee'}$ is dimension free (see Appendix \ref{sec:appendixA} for the explicit expression) and the only background length dependence that appears is as function of the ratio $x = \frac{A^2}{2R^2} = 1-\cos{\frac{2\pi}{N_A}}$. In the Hessian we have the background volume $V_{\sigma}$ given in \eqref{Volume}.

This scaling property of the Hessian motivates the replacement of the fluctuation variables $\lambda_e$ with rescaled variables 
\ba
\hat \lambda_e:= \tfrac{L_e}{\sqrt{6 V_\sigma}}\ \lambda_e \q .  
\ea

As expected $H^{\rm pr}$ has a null vector $\lambda_e\sim L_e$ that corresponds to the scaling symmetry discussed in section \ref{Sec:Diff}. For $M_{ee'}$ the corresponding null vector is given by $\hat \lambda_e \sim L_e^2$. In fact as $M^{\rm pr}$ does not depend on the background time edge length $T$, there is a second null vector, which corresponds to a scaling of the prism in time direction only. That is the null vector has only components for the $t$ and $d$ and $\tau$ as well as $\eta$ entries.  This additional null vector will lead to a null vector for the boundary theory that will correspond to (boundary) diffeomorphism symmetry in time direction. 

\subsubsection{Hessian for the full triangulation}\label{Sec:FH}

The Hessians for all the prisms can be added up to give the Hessian for the whole triangulation. To make it more transparent, we perform a lattice Fourier transform. Let us first consider a Fourier transform in angular direction only:
\ba
\hat \lambda(k,n)\,=\, \frac{1}{\sqrt{N_A}} \sum_s e^{ -i \frac{2\pi}{N_A}k\cdot s} \, \hat \lambda(s,n) \q ,\q\q   \hat\lambda(s,n)\,=\, \frac{1}{\sqrt{N_A}} \sum_k e^{ i \frac{2\pi}{N_A}k\cdot s} \, \hat\lambda(k,n) \q ,
\ea
where $k \in \{0,1,\dotsc, N_A-1\}$. We do not (angular) Fourier transform the time variables $\hat t(n)$ as they do not carry an angular dependence.  

The Fourier transform in time direction has to take into account the angular twist, which leads to an unusual periodic identification in time.  Let us parametrize the twist angle by $N_\gamma$ through 
\ba
\gamma\,=\, 2\pi \frac{N_\gamma}{N_A} \q .
\ea
Then the periodic identification of the variable is given by 
\ba
\hat\lambda(s,N_T)\,=\, \hat\lambda(s-N_\gamma,0) \q
\ea
which for the Fourier transformed variables translates as
\ba
\hat\lambda(k,N_T) \,=\, e^{-i \frac{2\pi N_\gamma}{N_A}k} \hat\lambda(k,0) \,=\, e^{-i \gamma k} \hat\lambda(k,0) \q .
\ea
Thus, the phase shifted variables
\ba
\hat\lambda'(k,n)\,=\, e^{i \gamma k \frac{ n}{N_T}}  \hat\lambda(k,n)
\ea
are periodic in the usual way, i.e. $\hat\lambda'(k,n+N_T)=\hat\lambda'(k,n)$. We therefore define the Fourier transform in time direction as
\ba
\hat\lambda(k,\nu)\,=\, \frac{1}{\sqrt{N_T}} \sum_n e^{ -i \frac{2\pi}{N_T} (\nu - \frac{\gamma}{2\pi}k  ) \cdot n} \, \hat\lambda(k,n)  \, ,\q\q    \hat\lambda(k,n)\,=\, \frac{1}{\sqrt{N_T}} \sum_\nu e^{ i \frac{2\pi}{N_T}(\nu - \frac{\gamma}{2\pi}k  ) \cdot n} \hat\lambda(k,\nu) \q,\q\q
\ea
with $\nu \in \{0,1,\cdots, N_T-1\}$. In the following we will use the shorthand notation
\ba
v:=\nu - \frac{\gamma}{2\pi}k  \q 
\ea
for the `twisted' time frequency.

The linearized action is then encoded as follows 
\ba\label{3.12a}
S^{(2)}\,\,=\,\, \frac{1}{2}\sum_{e,e'} \lambda_e H_{ee'} \lambda_{e'} \,\,=\,\, \frac{1}{2} \sum_{k,\nu}  (\hat\lambda(k,\nu))^{\rm t} \, \cdot \,\tilde M(k,\nu) \,\cdot \, (\hat\lambda(-k,-\nu))
\ea
where $(\hat \lambda(k,\nu))^{\rm t}$ stands for the transposed vector
\ba\label{3.13a}
(\hat\lambda(k,\nu))^{\rm t} \,=\, (\hat t(\nu),\hat r(k,\nu), \hat d(k,\nu), \hat \tau(k,\nu), \hat \alpha(l,\nu),\hat \eta(l,\nu)   ) \q .
\ea

The matrix $\tilde M(k,\nu)$ is given as 
\ba\label{Hess2}
\tilde M(k,\nu)&=&
{\scriptsize
\left(
\begin{matrix} 
0&-2x \sqrt{N_A} \delta_{k,0} &0 &0&\sqrt{N_A}\delta_{k,0}&0\\
-2x \sqrt{N_A} \delta_{k,0}& \Delta_k & -\Delta_k+2 x(1-\omega_v) &  \omega_v(\omega_k-1+2x)& ( \omega_v\omega_k-1) & \omega_v(1-\omega_k)\\ 
0&-\Delta_k+2 x(1-\omega^{-1}_v)& \Delta_k & (-1+\omega_k^{-1})& (\omega_v^{-1}-\omega_k)& (-1+\omega_k)\\
0& \omega^{-1}_v(-1+\omega_k^{-1}+2x)&(-1+\omega_k)&1& \frac{1}{2}(\omega_k- \omega_v^{-1})& -\omega_k\\
\sqrt{N_A}\delta_{k,0}&(-1+\omega_k^{-1}\omega_v^{-1}) & (\omega_v-\omega_k^{-1}) &  \frac{1}{2}(\omega^{-1}_k-\omega_v)&1& -\frac{1}{2}(1+\omega_v)\\
0&\omega_v^{-1}(1-\omega_k^{-1})&(-1+\omega_k^{-1})&-\omega_k^{-1}& -\frac{1}{2}(1+\omega^{-1}_v)&1
\end{matrix}
\right)
}
\nn\\
\ea 
where
\ba
\omega_k\,:=\, e^{i \frac{2\pi}{N_A} k} \,,\q\q \Delta_k=2- \omega_k-\omega_k^{-1} \, ,\nn\\
\omega_v\,:=\,e^{i \frac{2\pi}{N_T} v}\,,\q\q \Delta_v=2- \omega_v-\omega_v^{-1} \,.
\ea
Note that $\Delta_k$ and $\Delta_v$ are the eigenvalues of the Laplacians in the angular and time directions respectively.

From the discussion in section \ref{Sec:Diff} we expect that the matrix  $\tilde M(k,\nu)$ has a null vector for each independent bulk vertex displacement.  It turns out that these bulk vertex displacements only affect the (spatial) frequencies $k=0$ as well as $k=\pm 1$ (note that $k=-1$ is equivalent to $k=(N_A-1)$). To see that these vertex displacements do indeed lead to null vectors it is necessary to use the explicit value (\ref{t1}) for $x$ which amounts to 
\ba
x\,=\, \frac{1}{2} \Delta_{k=1}\,=\,1-\cos(\frac{2\pi}{N_A}) \q .
\ea
Remember that we have only bulk vertices along the central time axis. A displacement of  a bulk vertex in time direction, leads to a null vector  for $M_{k=0,\nu}$ given by \ba\label{null1}
(n_t)^{\rm t}(k=0,\nu)\, =\, (\frac{1}{\sqrt{N_A}}(1-\omega_v), 0,-\omega_v,0,0,0)
\ea
i.e. only the $\hat t(\nu)$ and the $\hat d(k=0,\nu)$ variables are affected (to linear order). See Appendix \ref{AppDiff} for an explanation of how to derive the null vectors geometrically. 
Then we have two independent displacements of a bulk vertex in the spatial hypersurface. They lead to null vectors for $M_{k=\pm 1,\nu}$ given by 
\ba
(n_r)^{\rm t}(k=\pm 1,\nu)\,=\, (0,\omega^{-1}_v,1,0,0,0)
\ea
i.e. only the $\hat r(k=\pm1,\nu)$ and the $\hat d(k=\pm 1,\nu)$ variables are affected.

These null vectors correspond exactly to the three independent bulk vertex displacements we expect to arise from diffeomorphism symmetry. They leave the boundary variables fixed and are thus proper gauge symmetries. As they only affect the $k=0$ and $k=\pm 1$ modes it will be much simpler to factor out these gauge symmetries as compared to the continuum treatment \cite{Barnich1}.

There are also null vectors of the Hessian which do affect the boundary variables. Thus we have for frequencies $|k|>0$ (and even keeping the value of $x$ general) the null vector 
\ba\label{null3}
(n_\tau)^{\rm t}(k,\nu)\,=\,  (0,0, 1,  \, 1-\omega_v, \,0  , \, 1-\omega_v\omega_k) \q .
\ea
Geometrically this null vector corresponds to  making the time slice $n$ smaller in time direction and the time slice $(n-1)$ larger by the same amount. This null vector will descend to a null vector for the boundary theory describing diffeomorphism symmetry in time direction. It is also clear that it comes from the scaling symmetry in time direction for the Hessian of the prism building block. 

The vector (\ref{null3}) is actually also null with respect to $\tilde M(k=0,\nu)$. Together with the null vector (\ref{null1}) it gives
\ba\label{null4}
(n_t+n_{\tau})^{\rm t}(k,\nu)=(  \frac{1}{\sqrt{N_A}}(1-\omega_v),0,1-\omega_v,1-\omega_v,0,1-\omega_v \omega_k)
\ea
and describes geometrically the same transformation as (\ref{null3}) for  the case $k=0$. 

Finally we still have the global scaling symmetry, which for $\tilde M(k=0,\nu=0)$ corresponds to a null vector
\ba
(n_{\rm sc})^{\rm t} (k=0,\nu=0)\,=\, ( \frac{T^2}{\sqrt{N_A}}, R^2, R^2+T^2,T^2, A^2, A^2+T^2) \q .
\ea


\section{Integrating out the bulk variables}\label{Sec:Int}

In this section we are going to integrate out the bulk variables. Let us first remark that so far it is not clear how to obtain the wished for result for the one--loop correction, which is encoded in the determinant of the Hessian for the bulk variables. Due to its dependence on the twist angle $\gamma$ we expect that the structure of the continuum result (\ref{Z1}) does require in particular a coupling of the time slices. However these appear  in the bulk Hessian only for the non--diagonal elements and are moreover multiplied with a factor $x$ which goes to zero either in the limit of large radius $R$ or in the continuum limit where $A$ goes to zero. It will turn out that integrating out e.g. the $\hat d$ variables, we obtain an action that is much more transparent. This is particularly relevant if one wants to understand the $\hat r$ variables as degrees of freedom which describe the position of the boundary in the embedding spacetime, as in a proposal of Carlip in \cite{carlip} for the AdS case.

\subsection{Treatment of gauge modes}

Let us start with the Fourier modes affected by the gauge symmetries. This will be important for the understanding of the structure of the one--loop correction.

We have seen that the Hessian for the bulk variables has three null eigenvectors per vertex, that is per mode number $\nu$. Those three null vectors result from the $k=0$ and $k=\pm 1$ Fourier modes respectively.

At $k=0$, we have $\omega_{k=0}=1$ and the  part of the matrix $\tilde M(k,\nu)$ describing the bulk variables $\hat t(\nu),\hat r(k=0,\nu), \hat d(k=0,\nu)$ is given by
\ba
\tilde M_{\rm bulk}(0,\nu)&=&\begin{pmatrix} 0& -2x \sqrt{N_A}& 0\\ -2x \sqrt{N_A}& 0& 2x(1-\omega_v)\\ 0& 2x(1 - \omega^{-1}_v)& 0\end{pmatrix} \q .
\ea
This matrix has one null vector. We proceed by integrating out the $\hat r(k=0,\nu), \hat d(k=0,\nu)$ variables. The corresponding sub--matrix has  two eigenvalues $\pm 2x \sqrt{\Delta_\nu}$. (Here $\Delta_\nu=\Delta_v$ with $v=\nu$ at $k=0$.) The negative sign of one of the eigenvalue reflects the conformal factor mode problem -- we analytically continue it to a positive sign. We then obtain
\ba\label{5.2a}
&&\!\!\! \int \!\! d \hat r(0,\nu) d\hat d(0,\nu)d \hat r(0,-\nu) d\hat d(0,-\nu) \, \times \nn\\ 
&&\q \exp\left( -\frac{1}{2} (\hat t(\nu),\hat r(0,\nu), \hat d(0,\nu)) \cdot \tilde M_{\rm bulk}(0,\nu) \cdot (\hat t(-\nu),\hat r(0,-\nu), \hat d(0,-\nu))^{\rm t}\right)
\,=\, \frac{4\pi^2}{4 x^2 \Delta_\nu}  \, , \q\q\q
\ea
that is a $\hat t$--independent result. This also holds if we consider in addition the boundary variables, and the $k=0$ part of the effective boundary action coincides with the result for generic frequencies given in (\ref{Hess4}). Note that here we integrate over the modes $(0,+\nu)$ and $(0,-\nu)$, thus if we take the product over the modes $\nu=0,\ldots, N_T-1$, the contribution of the $k=0$ modes to the partition function is in fact the square root of  (\ref{5.2a}).

~\\
At $k=\pm 1$ we have the bulk variables $\hat r(\pm 1,\nu),\hat d(\pm 1,\nu)$ whose dynamics is described by the following submatrix of the Hessian
\ba
\tilde M_{\rm bulk}(\pm 1,\nu)&=&
\begin{pmatrix} 2x& -2x\omega_v\\ -2x\omega^{-1}_v& 2x \end{pmatrix}\q .
\ea
It has one null vector (for each of the $k=\pm 1$ modes).  We integrate over the $\hat d(k=\pm 1,\nu)$ variable and are left with a $\hat r$--independent result. Also the $k=\pm 1$ part of the effective boundary action coincides with the general result given in (\ref{Hess4}). 
From the integration over $\hat d(k= 1,\nu)$ and $\hat d(k= -1,\nu)$ we obtain a factor 
\ba
\frac{2\pi}{2x} \q .
\ea

We now proceed by removing the remaining integration measure from the path integral. As explained in section \ref{Sec:Meas} 
 the measure that needs to be removed corresponds to the measure over the gauge orbit describing vertex translation invariance of the Regge action. That is for each vertex $p(n)$ on the axis of our solid torus we remove a measure factor 
 \be
 \frac{1}{(8\pi G\hbar)^{3/2}} \frac{1}{2\pi}dx^1(n) dx^2(n) dx^3(n)
 \ee
where $x^1(n),x^2(n),x^3(n)$ are the cartesian coordinates of the vertex $p(n)$. We thus need to translate these coordinates to our variables $t(\nu)$ and $r(\pm 1,\nu)$. Identifying $x^3$ with the $t$--direction we have
\ba
t(n)\,=\, x^3(n+1)-x^3(n)    \q \Rightarrow \q\q t(\nu) \,=\, (\omega_v -1) \, x^3(\nu)
\ea
which for the measure means 
\ba
\prod_n dx^3(n) \,=\, \prod_\nu \frac{1}{\sqrt{\Delta_\nu}} dt(\nu) \q .
\ea

We have furthermore the vertex displacement invariance in $x^1$ and $x^2$ direction (at fixed time step $n$). We identify the $x^1$ axis with the edge carrying the $r(s=0,n)$ variable. The displacement of the vertex $p(n)$ is then described by
\ba
(R+r(s,n))^2 \,=\, (R\cos(2\pi s/N_A)-\delta x^1(n))^2 + (R\sin(2\pi s /N_A)-\delta x^2(n))^2 
\ea
which to linear order in $r$ and $\delta x$ amounts to
\ba
r(s,n) \,\simeq\, -\delta x_1 \cos(2\pi s/N_A) -\delta x_2 \sin(2\pi s /N_A) \q .
\ea
Multiplying by $\omega_{k=\pm 1}^s$ and summing over $s$  we obtain
\ba
\begin{pmatrix}
r(k=+1,n) \\ r(k=-1,n)
\end{pmatrix}
\,\simeq\, \frac{\sqrt{N_A}}{2}
\begin{pmatrix}
-1&+i\\-1&-i
\end{pmatrix}
\begin{pmatrix}
\delta x^1(n) \\ \delta x^2(n)
\end{pmatrix} \q .
\ea
This leads to a Jacobian 
\ba
\prod_n dx^1(n) dx^2(n) \,=\,  \prod_\nu  \frac{2}{N_A} dr(+1,\nu) dr(-1,\nu) \q .
\ea
In summary the measure term corresponding to the gauge orbits is
\begin{equation}
\prod_n  \frac{dx^1(n) dx^2(n) dx^3(n)}{2\pi (8\pi G\hbar)^{3/2}} = \prod_\nu    \frac{1}{2\pi (8\pi G\hbar)^{3/2}}  \frac{1}{\sqrt{\Delta_\nu}}  \frac{2}{N_A} dr(+1,\nu) dr(-1,\nu)  dt(\nu)  \q .
\end{equation}
The usual Faddeev-Popov gauge-fixing gives essentially the same result.

\subsection{The physical modes}

We are thus left with the modes $k \in \{2, \dotsc, N_A-2\}, \nu \in \{0, \dotsc, N_T-1\}$ to integrate over.  Starting with the quadratic action (\ref{3.12a}) and  integrating out the $\hat d$ variables gives an effective action described by the following matrix
\ba\label{Hess3}
&&\tilde M_{r}(k,\nu) =\nn\\
&&{\scriptsize
\left(
\begin{matrix} 
2 x \Delta_v (1- \frac{2x}{\Delta_k})\,&\, (\Delta_k-2x)\frac{(\omega_k^{-1}-\omega_v)}{(1-\omega_k^{-1})}\,&\, (1-\omega^{-1}_v)(-1+ \omega_v\omega_k)  (1- \frac{2x}{\Delta_k})  \, &\,  (\Delta_k-2x)\frac{(-1+\omega_v)}{(1-\omega_k^{-1})}\\
(\Delta_k-2x)\frac{(\omega_k-\omega_v^{-1})}{(1-\omega_k)}\,&\, 0\, & \, \frac{1}{2} \frac{(1+\omega_k)(\omega_k-\omega_v^{-1})}{(1-\omega_k)}  \,&0\,  \\
(1-\omega_v)(-1+ \omega^{-1}_v\omega^{-1}_k)  (1- \frac{2x}{\Delta_k}) \, &\, \frac{1}{2} \frac{(1+\omega^{-1}_k)(\omega^{-1}_k-\omega_v)}{(1-\omega^{-1}_k)}\,&\,  \frac{ (\omega_k \omega_v+\omega^{-1}_k \omega^{-1}_v-\omega_k-\omega_k^{-1})}{\Delta_k}\,  & \, \frac{1}{2} \frac{(1- \omega_v)(1+\omega_k)}{(1-\omega_k)} \\
(\Delta_k-2x)\frac{(-1+\omega_v^{-1})}{(1-\omega_k)}\,&\,0\,&\, \frac{1}{2} \frac{(1- \omega^{-1}_v)(1+\omega^{-1}_k)}{(1-\omega^{-1}_k)}  & 0
\end{matrix}
\right),
}\nn\\
\ea 
for the fluctuations $(\hat r,\hat \tau,\hat \alpha, \hat \eta)$.

Integrating out in addition the $\hat r$ variables we get the effective action for the boundary fluctuations described by 
\ba\label{Hess4}
\tilde M_{\rm b}(k,\nu) &=&
\frac{1}{2x}  {\scriptsize
\left(
\begin{matrix} 
\frac{-\Delta_k \Delta_{kv}}{\Delta_v}    &  \frac{-\Delta_{kv}(1-\omega_k)}{(1-\omega_v)}    & \frac{ \Delta_k (\omega_k-\omega_v^{-1})}{(1-\omega_v^{-1})} \\
\cdots & -\Delta_{kv} & (1-\omega^{-1}_k)(1-\omega_k\omega_v) \\
\cdots & \cdots & -\Delta_k
\end{matrix}
\right)   
} 
+  {\scriptsize
\left( 
\begin{matrix} 
\frac{\Delta_{kv}}{\Delta_v} &             \frac{ (1+\omega_v)(\omega^{-1}_v-\omega_k)}{2 (1-\omega_v)} & - \frac{ (\omega_k-\omega_v^{-1})}{(1-\omega_v^{-1})} \\
\cdots & 1& -\frac{1}{2}(1+\omega_v) \\
\cdots & \cdots & 1
\end{matrix}
\right)
}\q\q
\ea
where the missing entries are found imposing that $\tilde M_{\rm b}(k,\nu)$ is a hermitian matrix. Moreover
\ba
\Delta_{kv}=2-\omega_k\omega_v-(\omega_k\omega_v)^{-1} \q .
\ea
Note that the $x^{-1}$ order part of the matrix will lead in the continuum limit to second order derivatives (counting inverse derivatives with a negative sign), whereas the $x^0$ order part gives rise to zeroth order derivatives.  
The $x^{-1}$ part of (\ref{Hess4}) is also invariant under time and angular (linearized) diffeomorphisms, whereas the $x^0$ order part is only invariant under time diffeomorphisms.

\subsection{The one--loop correction}\label{Sec:Loop}

Let us consider the one--loop determinant denoted $D$ in \eqref{PF2}. By definition, it is the determinant of the Hessian restricted to the bulk variables (equivalently for vanishing boundary fluctuations). For each mode $k\geq 2$, this Hessian is the two-by-two matrix 
\ba
\left(\begin{smallmatrix} \Delta_k& -\Delta_k + 2x(1-\omega_v)\\ -\Delta_k + 2x(1-\omega_v^{-1})& \Delta_k \end{smallmatrix}\right),
\ea
 where $\Delta_k$ is the eigenvalue of the Laplacian in the angular direction at frequency $k$.

Integrating out the variables $\hat d$ in the above matrix leads to a scalar, which is the diagonal $\hat r$ entry of $\tilde{M}_r(k, \nu)$ in \eqref{Hess3}, i.e.
\begin{equation}
2x \Delta_v \Bigl(1 - \frac{2x}{\Delta_k}\Bigr)\q .
\end{equation}
Before moving on and taking the product over the Fourier modes, let us comment on this quantity.

As the continuum result of \cite{Barnich1} depends on the twisting angle $\gamma$, we expect that it is essential to have a coupling between time slices in the (bulk) Hessian. This coupling appears in a rather weak form in (\ref{Hess2}): firstly only in the non--diagonal matrix entries, and secondly multiplied with the parameter $x$ which vanishes in the continuum limit.

Somewhat surprisingly, it turns out that after integration over the $\hat d$ variable (or the $\hat r$ variable which would have led to the same result for the remaining bulk part of the Hessian) the terms of zeroth order in $x$ vanish and one is left with first and second orders in $x$. 

Furthermore the non--local structure one expects from inverting the spatial Laplacian $\Delta_k$ (in the diagonal $\hat d$ component of $\tilde{M}(k, \nu)$ \eqref{Hess2}) is restricted to the second order term in $x$. We also obtain from the partial integration of the $\hat d$ variables the Laplacian $\Delta_v$ in time direction. This Laplacian will be responsible for the $\gamma$--dependent part of the one--loop correction.

As discussed in section \ref{Sec:Meas} the initial measure for the path integral is
\ba\label{5.16}
&& \prod_{e \in {\rm bdry}} \frac{L_e^{1/2}}{(8\pi G\hbar)^{1/4}}    \prod_{e \in {\rm bulk}} \frac{L_e}{(8\pi G\hbar)^{1/2}}  d\lambda_e  \prod_{\sigma} \frac{1}{(12\pi V_{\sigma})^{1/2}} \nn\\ &&\q\q\q\q
\,=\,
\left[\prod_{e \in {\rm bdry}} \frac{L_e^{1/2}}{(8\pi G\hbar)^{1/4}} \right] \,\, \frac{(6V_{\sigma})^{N_AN_T + N_T/2}}{(12\pi V_{\sigma})^{\tfrac{3}{2} N_A N_T}} \frac{1}{(8\pi G\hbar)^{N_A N_T + N_T/2}}  \prod_{e \in {\rm bulk}} d\hat \lambda_e \q .\q\q\q
\ea

We will now collect all factors arising from the integration over the various bulk variables. The interesting $\gamma$--dependent part will come from integrating out the $\hat r$ variables, and arise there from the determinant of the entry $\Delta_v$ in the effective Hessian (where one already integrated out the $\hat d$ variables). 

From the integration over the modes $(k=0,\nu)$ and $(k=\pm 1,\nu)$ and from taking into account the measure over the gauge orbits we obtain a factor
\ba\label{5.17}
(8\pi G\hbar)^{\tfrac{7}{2} N_T} \left(\frac{2\pi}{2x}\right)^{2 N_T} (\pi N_A)^{N_T} \frac{ (6V_{\sigma})^{\tfrac{3}{2}N_T}}{R^{2N_T} T^{N_T}}
\ea
Note in particular that a $\sqrt{\Delta_\nu}$ contribution from the $k=0$ mode cancels with the contribution from the gauge orbit measure.

Next we consider the integration over the $\hat d$ variables for the modes $k=2,\ldots, N_A-2$ and $\nu=0,\ldots, N_T-1$. The integration of
\ba
\exp\left(-\frac{1}{2 (8\pi G \hbar)} \hat d(k,\nu)\, \Delta_k \, \hat d(-k,-\nu)  \right)
\ea
leads to a factor
\ba\label{5.19a}
(8\pi G \hbar)^{ \tfrac{N_T(N_A-3)}{2} } \prod_{\nu=0}^{N_T-1} \prod_{k=2}^{N_A-2} \frac{(2\pi)^{1/2}}{\sqrt{\Delta_k}} \q .
\ea
Using the result
\ba
\prod_{k=1}^{N_A-1} \Delta_k \,=\, N_A^2 \q \Rightarrow \q \prod_{k=2}^{N_A-2} \Delta_k \,=\, \frac{N_A^2}{\Delta_{k=1}^2}\,=\, \frac{N_A^2}{(2x)^2}
\ea
we can rewrite this factor (\ref{5.19a}) into
\ba\label{5.21}
(8\pi G \hbar)^{ \tfrac{N_T(N_A-3)}{2} }  (2\pi)^{ \tfrac{N_T(N_A-3)}{2} }  \left(\frac{2x}{N_A}\right)^{N_T}  \q .
\ea

Finally we consider the integration over the $\hat r$ variables, again for the modes $k=2,\ldots, N_A-2$ and $\nu=0,\ldots, N_T-1$.
To this end we take the product over the modes of the $(\hat r,\hat r)$ component in (\ref{Hess3}):
\ba
\prod_{k=2}^{N_A-2} \prod_{\nu=0}^{N_T-1}  \,2 x \, \Delta_v \left(1- \frac{2x}{\Delta_k}\right)   &=& (2x)^{N_T(N_A-3)} \left( \prod_{k=2}^{N_A-2} \left(1- \frac{2x}{\Delta_k}\right)^{N_T}\right)
\left(
\prod_{k=2}^{N_A-2} \prod_{\nu=0}^{N_T-1}  \Delta_v
\right) \, .\q
\ea
For the product involving the time frequencies we obtain
\begin{equation}
\prod_{\nu=0}^{N_T-1} \Delta_v = \prod_{\nu=0}^{N_T-1}  \left(2 - \exp\left( i \frac{2\pi}{N_T} (\nu - \gamma \frac{k}{2\pi} )\right)  -  \exp\left( -i \frac{2\pi}{N_T} (\nu - \gamma \frac{k}{2\pi} )\right) \right) = 
2-2\cos(\gamma k ) \q .
\end{equation}

For the other factor we remember that $2x=\Delta_{k=1}$ and use the following identity
\ba
\prod_{k=2}^{N_A-2} \left(1- \frac{2 - a-a^{-1}}{2-a^k - a^{-k}} \right) &=&\frac{ 1-a^{N_A-1} }{(1+a)(1-a^{N_A-2})}
\ea
which gives applied to our case (with $\Delta_k=2-a^k-a^{-k}$ and $a^{N_A}=1$)
\ba
\prod_{k=2}^{N_A-2} \left(  1- \frac{ 2x }{\Delta_k} \right) &=& \frac{1}{4-2x}  \q .
\ea
In summary we obtain
\ba
\prod_{k=2}^{N_A-2} \prod_{\nu=0}^{N_T-1}  \,2 x \, \Delta_v \left(1- \frac{2x}{\Delta_k}\right)   &=& \frac{ (2x)^{N_T(N_A-3)} }{ (4-2x)^{N_T}}\,\, \prod_{k=2}^{N_A-2}  (2-2\cos(\gamma k )) \q .
\ea
This leads to the following factor from the integration over the $\hat r$ modes
\ba\label{5.27}
 (8\pi G \hbar)^{ \tfrac{N_T(N_A-3)}{2} }      (2 \pi)^{ \tfrac{N_T(N_A-3)}{2} }    \frac{(4-2x)^{N_T/2}}{(2x)^{\tfrac{1}{2} N_T(N_A-3)} }
    \left[ \prod_{k=2}^{N_A-2}  (2-2\cos(\gamma k ))\right]^{-1/2}   \q .
     \ea

Collecting all factors in (\ref{5.16},\ref{5.17},\ref{5.21},\ref{5.27}) and using $6V_\sigma=\frac{1}{2} ATR\sqrt{4-2x}$ as well as $2x=A^2/R^2$ we obtain
\ba\label{DON}
&& D = 2^{-N_T} (2\pi)^{-\frac{N_T N_A}{2}} 
\left(\frac{A}{R}\right)^{-N_T(N_A-1)}  (ART)^{-\frac{N_T N_A}{2}} A^{ 2 N_T} T^{N_T}
 \left(4-\frac{A^2}{R^2}\right)^{-\frac{N_T N_A}{4}+ \frac{3N_T}{2}}  \,\, \nn\\&&\q\q\q\q\q\q\q\q \times \left[ \prod_{e \in {\rm bdry}} \frac{L_e^{1/2}}{(8\pi G\hbar)^{1/4}} \right]  
\left[ 
\prod_{k=2}^{(N_A-1)/2} \frac{1}{|1-q^k|^2}
\right]
\ea
where $q=\exp(i\gamma)$ and we assumed that $N_A$ is odd for simplicity. Here one can use the relation $A^2/(2R^2)=1-\cos(2\pi/N_A)$ to eliminate either $R$ or $A$.

The terms which are independent of $\gamma$ ensure the consistent gluing of partition functions along discrete boundaries.
 
The interesting part is the $\gamma$--dependent term, $\prod_{k=2}^{(N_A-1)/2} |1-q^k|^{-2}$, which in the limit $N_A \rightarrow \infty$ reproduces the one--loop correction in the continuum (\ref{Z1}). Our discrete approach further offers the following insights.
\begin{itemize}
\item We see that the one--loop correction is caused by degrees of freedom (the $\hat r$ variables) that connect to the boundary.
\item The product over $k$ results from a product over angular Fourier modes.
\item That the product starts with $k=2$ is explained by the diffeomorphism invariance of the action. This  invariance of the action means that the vertices at the axis can be (infinitesimally)  displaced without changing the value of the Hamilton--Jacobi functional. This vertex displacement involves only length perturbations at $k=0$ and $k=\pm 1$, explaining that the one--loop determinant starts with $k=2$.
\end{itemize}

Let us comment that without the regularizing shift $q=\exp(i\gamma) \to q=\exp(i\gamma -\varepsilon)$, the continuum result (\ref{Z1}) is explicitly divergent for all rational angles of the form $\gamma = 2\pi p/p'$ with $p,p' \in \mathbb{N}$ and $\text{gcd}(p,p')=1$. The reason is that one encounters a factor $1/0$ in the product over $k$ when $k$ is a multiple of $p'$. In the discrete case we can only deal with rational angles. The minimal $N_A$ we can choose to accommodate such a twisting angle is $N_A=p'$. In this case the product in (\ref{DON}) is finite without regularizing shift. However going to a refinement that also allows the same twisting angle, e.g. $N_A=2p'$ (and considering the slight generalization of (\ref{DON}) for even $N_A$) or $N_A=3p'$ we also get a divergent factor $1/0$.  Thus we either have to also introduce a regularizing shift, or adjust $N_A$ to the twisting angle and choose $N_A=p'$.

\subsection{ The effective boundary action} \label{Sec:Eff}

Considering the partition function with a boundary at finite radius (encoded in the zeroth order boundary data), we can also allow arbitrary boundary fluctuations and consider the Hamilton--Jacobi functional, that is the action evaluated on the solution, as a function of these boundary data. The Hamilton--Jacobi function encodes all information about the dynamics of the system and furthermore represents the classical approximation to the partition function. Furthermore we will see that the (boundary diffeomorphism invariant part of the) Hamilton--Jacobi functional can be described by a dual field theory, that will {\it a posteriori} explain the one--loop result. 

We evaluated the zeroth order part of the action in (\ref{S0R}) as $S_{HJ}^{(0)}=\frac{-\beta}{4G}$.  We found that the action has also a first order part
\ba \label{HJ1}
8 \pi G \, S_{HJ}^{(1)}\,=\,-\frac{2\pi }{N_A} \sum_{s,n} \tau(s,n)   \q .
\ea
We will consider the Hamilton--Jacobi functional for boundary data that are induced by infinitesimal diffeomorphisms.\footnote{Note that we consider only the change to first order in the boundary (length) variables and thus treat  the perturbation parameter as defined via the expansion of the boundary data.  Alternatively one could consider how a diffeomorphism of the background solution changes the boundary data including second order terms and include these into the first order action. This would lead to a second order contribution.} 
In the discrete this is equivalent to infinitesimal displacements of the boundary vertices, 
as discussed in the Appendix \ref{AppDiff}. There are three directions in which a vertex can be displaced, in time, angular and radial direction. To first order a vertex displacement in angular and radial direction does not change the edge lengths in time direction. For vertices moved along the time direction by an amount $\chi(s, n)$, one gets
\begin{equation} \label{BdryTimeDiff}
\tau(s, n) = \chi(s, n) - \chi(s, n+1).
\end{equation}
Since each $\chi(s, n)$ appears twice in the variation $S_{HJ}$, once with a positive sign and once with a negative sign, we have the discrete equivalent of a total divergence term.  
Thus $S_{HJ}^{(1)}$ is vanishing for all possible first order displacements of boundary vertices.

~\\

The second order part of the Hamilton--Jacobi functional for the linearized theory is encoded in the matrix $\tilde M_{\rm b}(k,\nu)$ given in (\ref{Hess4}), that is  
\ba\label{3.12b}
8\pi G\, \,S^{(2)}_{HJ}\,\,=\,\, \frac{1}{2} \sum_{k,\nu}  (\hat \lambda_{\rm b}(k,\nu))^{\rm t} \, \cdot \,\tilde M_{\rm b}(k,\nu) \,\cdot \, (\hat \lambda_{\rm b}(-k,-\nu))
\ea
where $(\hat \lambda_{\rm b}(k,\nu))^{\rm t}$ stands for
\ba\label{3.13b}
(\hat \lambda_{\rm b}(k,\nu))^{\rm t} \,=\, ( \hat \tau(k,\nu), \hat \alpha(k,\nu),\hat \eta(k,\nu)   )
\ea
with $\hat\cdot$ indicating the rescaling $\lambda_e \rightarrow \hat \lambda_e = \frac{L_e}{\sqrt{6 V_{\sigma}}} \lambda_e$ of the fluctuation variables.

We aim at comparing the continuum limit of the discrete results with results obtained directly from the continuum theory. To this end we should match the variables we are using  ($\lambda$ or $\hat \lambda$) to metric fluctuations.

The relation between the (boundary) length fluctuations $\lambda_{\rm b}$ and the metric fluctuations $\delta h_{ab}$ are determined by 
\ba\label{htrafo1}
(h_{a b} +\delta h_{ab}) e^a_\tau e^b_\tau &=& h_{\tau\tau}+\delta h_{\tau\tau}  \,=\, (T+\tau)^2 \,=\, T^2+2T \tau + {\cal O}(\tau^2) \nn\\
(h_{a b} +\delta h_{ab})  e^a_\alpha e^b_\alpha &=& h_{\alpha\alpha}+\delta h_{\alpha\alpha}  \,=\, (A+\alpha)^2 \,=\, A^2+2A \alpha + {\cal O}(\alpha^2)\nn\\
(h_{a b} +\delta h_{ab})  (e^a_\tau+e^a_\alpha) ( e^b_\tau+  e^b_\alpha ) &=& h_{\alpha\alpha}+ h_{\tau\tau} + 2 h_{\tau\alpha}  +\delta h_{\alpha\alpha}+ \delta h_{\tau\tau} +\delta 2 h_{\tau\alpha}\nn\\ &=& (\sqrt{A^2+T^2}+\eta)^2 \,=\, A^2+T^2+2\sqrt{A^2+T^2} \eta + {\cal O}(\eta^2)  \,.\nn\\
\ea
This fixes the background boundary metric (with respect to basis vectors $(e^a_\tau,e^a_\alpha)$) to $h_{ab}={\rm{diag}}(T^2,A^2)$. Taking into account the rescaling $\hat \lambda = \frac{L_e}{\sqrt{6 V_\sigma}} \lambda_e$ we have the following transformation between rescaled length and metric perturbations to first order:
\ba\label{htrafo2}
\left(
\begin{matrix}
\delta h_{\tau\tau}\\
\delta h_{\alpha\alpha}\\
\delta h_{\tau\alpha}
\end{matrix}
\right)
&=&
\sqrt{6 V_\sigma}
\left(
\begin{matrix}
2&0&0\\
0&2&0\\
-1&-1&1
\end{matrix}
\right)
\left(
\begin{matrix}
\hat \tau \\
\hat \alpha \\
\hat \eta
\end{matrix}
\right)  +{\cal O}(\lambda^2) .
\ea
However the full transformation between length and metric variables is non--linear and -- as we have a non-vanishing first order action -- we will need the $\tau$ variables expressed as function of the metric variables to second order
\ba\label{htrafo3}
\tau&=& \frac{1}{2T} \delta h_{\tau\tau} \,-\, \frac{1}{8 T^3} (\delta h_{\tau \tau})^2 + {\cal O}( (\delta h_{\tau \tau})^3)  \q .
\ea



As we are dealing with different variables, which transform non--linearly, we need to be careful with treating the order of the perturbations correctly.  
For instance we will have two contributions for the effective action in metric variables: (a) One part is given by the second order part of the effective action in length variables transformed via the linear transformation (\ref{htrafo2}) to the metric perturbations. (b) The second part is given by the first order action in length variables, with the length variables expressed in metric perturbations to second order, as in (\ref{htrafo3}).
Both contributions give 
\ba\label{3.12}
8\pi G\, \,S^{(2)}_{HJ}\,\,=\,\, \frac{1}{2} \sum_{k,\nu}  (\delta h (k,\nu))^{\rm t} \, \cdot \,\tilde M^{\rm h}_{\rm b}(k,\nu) \,\cdot \, (\delta h (-k,-\nu))
\ea
with
\ba\label{Hess5}
\tilde M^{\rm h}_{\rm b}(k,\nu) &=&
-\frac{1}{8x \cdot(6 V_{\sigma})}  {\scriptsize
\left(
\begin{matrix} 
\frac{\Delta^2_k }{\Delta_v}    & \Delta_k   &  2  \frac{ \Delta_k   (1-\omega_k)}{(1-\omega_v^{-1})} \\
\cdots & \Delta_{v} &         2(1-\omega_k)(1-\omega_v) \\
\cdots & \cdots & 4 \Delta_k
\end{matrix}
\right)   
} 
+
\frac{1}{8\cdot(6 V_{\sigma})}   {\scriptsize
\left( 
\begin{matrix} 
2\frac{\Delta_{k}}{\Delta_v} & (1-\omega_k) &    4  \frac{    (1-\omega_k)}{(1-\omega_v^{-1})}     \\
\cdots & \Delta_v &2(1-\omega_v) \\
\cdots & \cdots & 8
\end{matrix}
\right)
} \,\, \nn\\
&&+ \frac{\pi}{8N_A T^3}
 {\scriptsize
\left(
\begin{matrix} 
1   & 0&  0 \\
\cdots & 0 &      0 \\
\cdots & \cdots & 0
\end{matrix}
\right)   
}   \q .
\ea
Note that the term originating from the first order action in length variables breaks the scaling property in $T$ direction, and will indeed be the only one braking the invariance under infinitesimal time diffeomorphisms.

The $x^{-1}$ order term of (\ref{Hess5}) will again lead to second order derivatives in the continuum limit. This part is also invariant under both the angular and the time (boundary) diffeomorphisms (by that we mean linearized diffeomorphisms, as in the rest of the paper). The second term is not invariant under angular diffeomorphisms  but is invariant under time diffeomorphisms. The third term, which we obtained from the first order result in length variables, is not invariant under time diffeomorphisms but is invariant under angular diffeomorphisms. The $x^0$ part includes terms with zero order derivatives (eg. the $(\tau\tau,\tau\tau)$ or $(\tau\tau,\tau\alpha)$ component) but also terms like $(1-\omega_k)$ (the $(\tau\tau,\alpha\alpha)$ component) or $\Delta_v$ (the $(\alpha\alpha,\alpha\alpha)$ component), which seem to lead to higher order derivatives. Note however that those terms  -- characterized by having at least one index being equal to $\alpha\alpha$ --  will actually vanish in the continuum limit considered here, so that the $x^0$ part will still  give rise to only zeroth order derivative terms. Thus for a perturbation whose only non--vanishing component is $\delta h_{\alpha\alpha}$, only terms from the $x^{-1}$ part of the matrix (\ref{Hess5}) survive in the continuum limit. This applies to the metric perturbations arising from a radial diffeomorphism (after having taken the continuum limit). It thus turns out that the first term in (\ref{Hess5}) is the only one relevant for radial diffeomorphisms, the second term sees the angular diffeomorphisms and the last term the diffeomorphisms in time direction. 

At this point one might wonder why the (second order) effective boundary action is not invariant under (first order) boundary diffeomorphisms. The answer is that we are considering here linearized diffeomorphisms for the second order term of the effective boundary action. The second order action would indeed be invariant to linear order, if the linear order of the action would be identically vanishing.  However the effective  action has also a non--vanishing first order term, for which one would need to consider diffeomorphisms to second order. These terms  cancel the terms resulting from the second order term of the action contracted with a metric perturbation from a linearized boundary diffeomorphism.

\subsection{The continuum limit}\label{Sec:Climit}

We can now take the continuum limit resulting in a continuum second order effective boundary action. This result will be confirmed later in Section \ref{Sec:Cont} by a direct computation in the continuum. 

We take the continuum limit by setting 
\ba\label{cl1}
A\,=\, \varepsilon A_0 \, \q,\q\q T\,=\, \varepsilon T_0
\ea
and considering $\varepsilon\ll 1$. This implies for the various other quantities appearing in (\ref{3.12})
\ba\label{cl2}
x&=& \frac{A^2}{2R^2}\,=\,  \varepsilon^2 \frac{A_0^2}{2R^2} \, ,\nn\\
N_T&=& \varepsilon^{-1} \frac{\beta}{T_0} \,  ,\nn\\
N_A&=& 2\pi  \arccos^{-1}\left(1-\frac{A^2}{2R^2}\right) \,\,=\,\, \varepsilon^{-1}\frac{2\pi R}{A_0} + {\cal O}(\varepsilon^0)\, , \nn\\
\omega_v &=& \exp( \frac{2\pi i }{N_T} v) \,\,=\,\, 1+   \varepsilon \frac{2\pi i T_0}{\beta} v + {\cal O}(\varepsilon^2) \,=:\, 1+   \varepsilon  i \hat v +{\cal O}(\varepsilon^2)\, ,\nn\\
\omega_k &=& \exp( \frac{2\pi i }{N_A} k) \,\,=\,\, 1+   \varepsilon \frac{ i A_0}{R} k +{\cal O}(\varepsilon^2)\,=:\,1+ \varepsilon i \hat k +\,{\cal O}(\varepsilon^2) ,\nn\\  
6 V_{\sigma} &=& ATR\sqrt{ 1-\frac{x}{2}}\,\,=\,\, \varepsilon^2 A_0T_0 R + {\cal O}(\varepsilon^3)   ,
\ea
where we defined rescaled wave vectors $\hat k$ and $\hat v$. We obtain for the matrix (\ref{Hess5})
\ba\label{Hesslimit}
\tilde M^{\rm h}_{\rm b}(k,\nu) &=&
-\frac{ \varepsilon^{-2} \,R}{4 A_0^3 T_0}  
\left(
\begin{matrix} 
\frac{\hat k^4}{\hat v^2}   & \hat k^2& -2 \frac{\hat k^3}{\hat v}\\
\cdots &   \hat v^2 &      -2 \hat k \hat v \\
\cdots & \cdots &  4 \hat k^2
\end{matrix}
\right)   
+
\frac{\varepsilon^{-2}}{4 A_0T_0R} 
\left( 
\begin{matrix} 
\frac{\hat k^2}{\hat v^2} &0&    -2 \hat{k}{\hat v}    \\
\cdots & 0 &0 \\
\cdots & \cdots & 4
\end{matrix}
\right)
 \,\, \nn\\
&&+ \frac{\varepsilon^{-2}\,A_0}{4  R T_0^3}
\left(
\begin{matrix} 
1   & 0&  0 \\
\cdots & 0 &      0 \\
\cdots & \cdots & 0
\end{matrix}
\right)   
   \q + {\cal O}(\varepsilon^{-1}) .
\ea

We aim to compare this result to a calculation performed directly in the continuum. There we will consider the action evaluated on perturbations of the flat metric induced by infinitesimal diffeomorphisms. We therefore will also evaluate in the discrete the Hamilton--Jacobi functional encoded in (\ref{Hesslimit}) on boundary data that describe such infinitesimal deformations. 
In the discrete these are given by vertex displacements and -- as derived in Appendix \ref{AppDiff} --  the vertex displacements along the boundary, that is in time and angular direction, lead to the following description in terms of metric perturbations $\delta h$,
\ba\label{bdrydiffeosA}
(n^{\rm h}_{{\rm b},\tau})^{\rm t}(k,\nu) &=&-\, (2(1-\omega_v) ,\,\, 0,\,\,  \omega_v(1-\omega_k) )  \,  \,\,X_\tau(k,\nu) \q , \nn\\
(n^{\rm h}_{{\rm b},\alpha})^{\rm t}(k,\nu) &=& \, (0 ,\,\, 2(1-\omega_k),\,\,  \omega_k(1-\omega_v) )  \,  \,\,X_\alpha(k,\nu)  \q . \q\q
\ea
Here $X_\tau,X_\alpha$ give the distance between old and new vertex positions. In the continuum limit defined by (\ref{cl1},\ref{cl2}), together with $X_\tau =\varepsilon X_\tau^0$ and $X_\alpha =\varepsilon X_\alpha^0$, we have
\ba\label{diffeoB}
(n^{\rm h}_{{\rm b},\tau})^{\rm t}(k,\nu) &=&\varepsilon^2 \,\,(2i\hat v , 0, i\hat k )  \,\, X^0_\tau(k,\nu)  \,\,+\,\,{\cal O}(\varepsilon^3)\q , \nn\\ 
(n^{\rm h}_{{\rm b},\alpha})^{\rm t}(k,\nu) &=&  \varepsilon^2 \,\, (0 ,\,\, 2i\hat k,\,\,  i\hat v )  \,  \,\,X^0_\alpha(k,\nu) \,\,+\,\,{\cal O}(\varepsilon^3) \q .
\ea
For these vectors we obtain the following action contributions
\ba\label{5.43}
8\pi G  S^{(2)}[ (n^{\rm h}_{{\rm b},\tau}) ]&=&\frac{1}{2} \sum_{k,\nu}  (n^{\rm h}_{{\rm b},\tau}(k,\nu))^{\rm t} \, \cdot \,\tilde M^{\rm h}_{\rm b}(k,\nu) \,\cdot \, (n^{\rm h}_{{\rm b},\tau}(-k,-\nu)) 
\nn\\ &=& \frac{1}{2}\sum_{k,\nu}  \varepsilon^2 \frac{A_0}{RT^3_0} \hat v^2  \,  X^0_\tau(k,\nu) X^0_\tau(-k,-\nu)  \,\,+\,\,{\cal O}(\varepsilon^3)\q , \nn\\
8\pi G  S^{(2)}[ (n^{\rm h}_{{\rm b},\alpha}) ]&=& \frac{1}{2} \sum_{k,\nu}  (n^{\rm h}_{{\rm b},\alpha}(k,\nu))^{\rm t} \, \cdot \,\tilde M^{\rm h}_{\rm b}(k,\nu) \,\cdot \, (n^{\rm h}_{{\rm b},\alpha}(-k,-\nu))\nn\\ &=& \frac{1}{2}\sum_{k,\nu}  \varepsilon^2 \frac{1}{RA_0T_0} \hat v^2 \,  X^0_\alpha(k,\nu) X^0_\alpha(-k,-\nu)  \,\,+\,\,{\cal O}(\varepsilon^3) \q . \q\q
\ea

We can furthermore consider an infinitesimal displacement of boundary vertices in radial direction, which is described by (see appendix)
\ba
(n^{\rm h}_{{\rm b},r })^{\rm t}(k,\nu) &=& A \sin( \tfrac{\pi}{N_A})\,(0 ,2( 1+\omega_k), (1-\omega_v^{-1} )\omega_v\omega_k) \, X_r(k,\nu) 
\ea
with $X_r$ giving the amount of the radial (outward) displacement. In the continuum limit (here we do not put $X_r= \varepsilon X_r^0$, as we do not take a continuum limit in $r$ direction) we have
\ba\label{diffeoR}
(n^{\rm h}_{{\rm b},r })^{\rm t}(k,\nu) &=& \varepsilon^2 \frac{A^2_0}{R} \, (0 ,2, 0) \, X_r(k,\nu) \,\,+\,\,{\cal O}(\varepsilon^3) \q .
\ea
 This leads to an action contribution
\ba\label{5.46}
8\pi G  S^{(2)}[ (n^{\rm h}_{{\rm b},r}) ]&=&\frac{1}{2} \sum_{k,\nu}  (n^{\rm h}_{{\rm b},r}(k,\nu))^{\rm t} \, \cdot \,\tilde M^{\rm h}_{\rm b}(k,\nu) \,\cdot \, (n^{\rm h}_{{\rm b},r}(-k,-\nu)) 
\nn\\ &=& -\frac{1}{2}\sum_{k,\nu}  \varepsilon^2 \frac{ A_0 }{R T_0} \hat v^2  \,  X_r(k,\nu) X_r(-k,-\nu)  \,\,+\,\,{\cal O}(\varepsilon^3) \q .
\ea

In summary the effective boundary action evaluated on  metric perturbations induced by diffeomorphisms has quite a simple form. Apart from scaling factors all three kinds of deformation are just subject to a Laplacian in time direction. The radial diffeomorphisms come with the opposite sign as compared to the boundary diffeomorphisms. Note also that the diffeomorphisms in the various directions are orthogonal to each other with respect to the quadratic form defined by the second order action. This follows from the partitioning of the matrix $\tilde M^{\rm h}_{\rm b}$ into three terms, where (in the continuum limit) each of the term is only non--vanishing on a diffeomorphism in either radial, angular or time direction.

\section{Boundary action and Liouville field theory} \label{Sec:L}

For the one--loop correction the terms surviving the continuum limit are those of lowest order in the parameter $x$. As we have seen the contributions to the effective boundary action split also in a part of order $x^{-1}$ and a part of order $x^{0}$. The part with order $x^{-1}$ leads in the continuum limit to second order derivatives. (The second order derivatives that we obtain for the boundary diffeomorphisms are just due the fact that we did not normalize the corresponding vectors.) This part is also invariant under the boundary diffeomorphisms, and hence is only sensitive to the \emph{spacetime} diffeomorphisms with displacement vector normal to the boundary.

We might thus look for a dual  theory of a scalar field on the boundary  coupled to  boundary metric variables, such that integrating out this scalar field will regain the (second order in derivatives part of the) effective boundary action in (\ref{Hess4}). Indeed such a dual field theory, given by a Liouville field theory, can be constructed in the AdS case \cite{skenderis}. Carlip \cite{carlip} derived such an action by arguing that the boundary effective action should result from the breaking of the spacetime diffeomorphism symmetry by the presence of the boundary. The scalar field encodes the radial distance and is thus related to the variables $r$ in our linearized Regge action, after integrating out the diagonal variables $d$. Thus $\tilde M_r$ in (\ref{Hess3}) is a candidate for such an action.  However the correspondence is spoiled by $\tilde M_r$ violating invariance under the boundary diffeomorphisms.   Also a discretization of Liouville theory linearized around a vanishing field should have vanishing entries for the entries quadratic in the boundary length fluctuations, i.e. of type $\lambda_{\rm b} \lambda'_{\rm b}$. This is not the case for (\ref{Hess3}). One can also look for a field transformation (adding to the $r$ field some linear combination of the boundary length perturbations) such that these entries do vanish. However the most one can accomplish is that all the terms quadratic in boundary length fluctuations do vanish except the $\hat \alpha(k,\nu) \hat \alpha(-k,-\nu)$ terms.  

We will indeed manage to match the $\lambda_r \lambda_{\rm b}$ terms of lowest order in $x$  with  a field of Liouville type\footnote{More precisely our ansatz for the action of the field will be the action of a free scalar field coupled to the boundary curvature as external current as for the Liouville action.}, thus confirming the interpretation of the $r$--variables as an Liouville type field, whose integration gives the (boundary diffeomorphism invariant part of the) effective boundary action.

To show this we discretize the  Liouville action (without a potential term), after which we will integrate out the Liouville field using the linearized field equations. We can then compare the resulting boundary action to the result (\ref{Hess4}).

We consider a Liouville action without potential given by
\ba\label{Lfield1}
S_L &=& \int d^2 y  \sqrt{h} \left(    h^{ab} \varphi \partial_a\partial_b \varphi  +  c\,\,  {}^{2D}\! R \, \varphi \right)
\ea
with $c$ defining the coupling of the 2D curvature to the scalar field $\varphi$.  We discretize this action for a 2D regular triangulation of the torus (coinciding with the boundary triangulation of our 3D spacetime) and then linearize the result.

The scalar field $\varphi(y)$ will be turned into a field $\varphi_{p}$ associated to the vertices $p$ of the triangulation. The kinetic term of the scalar field is discretized using the so--called cotangent rule for the discretization of the Laplacian (see e.g. \cite{sorkin}) on a triangulation
\ba
 \int d^2 y  \sqrt{h}   h^{ab} \varphi \partial_a\partial_b \varphi \q\q \rightarrow \q\q \sum_{\triangle} \sum_{e \in \triangle} \tfrac{1}{2}\cot(\alpha^{\triangle}_{\overline{e}}) ( \varphi_{s(e)} - \varphi_{t(e)})^2
\ea  
where the sum is over the triangles $\triangle$ and $\alpha^{\triangle}_{\overline{e}}$ denotes the interior angle at the vertex $p$ opposite the edge $e$ in $\triangle$, and $s(e),t(e)$ are the source and target vertex of $e$ respectively.  For a rectangle built from two triangles we obtain an action contribution
\ba
S_{K,\rm{rec}}&=& \frac{T}{2A}\left( (\varphi_1-\varphi_2)^2+  (\varphi_3-\varphi_4)^2\right) +  \frac{A}{2T}\left( (\varphi_1-\varphi_3)^2+  (\varphi_2-\varphi_4)^2\right)  \q .
\ea
Here the edges between vertices $p=1,2$ and $p=3,4$ have length $A$ and the edges between $p=1,3$ and $p=2,4$ are of length $T$.

\begin{figure}[htb]
\includegraphics[scale=0.5]{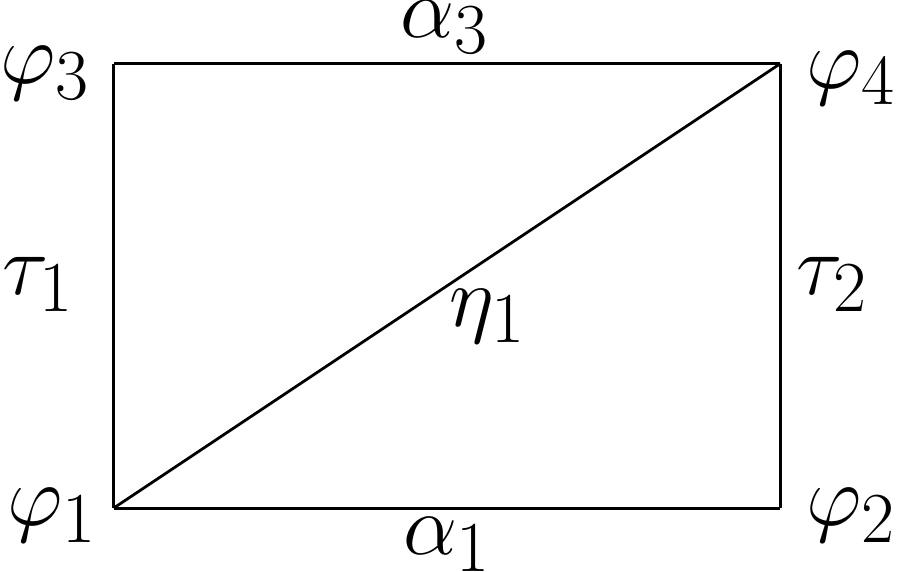}
\caption{\small  The Liouville field and length fluctuation variables for a boundary rectangle.}\label{fig_Li}
\end{figure}

The 2D curvature is discretized as
\ba
\int d^2 y  \sqrt{h} R  \, \varphi \q \q \rightarrow \q \q \sum_{p }  \epsilon_{p}\, \varphi_p
\ea
where $\epsilon_p\,=\, 2\pi - \sum_{\triangle} \alpha^{\triangle}_p$ is the deficit angle at the vertex $p$ of the triangulation.

Expanding this discrete action around the $\varphi=0$ configuration with vanishing deficit angles gives
\ba
 \sum_{p }  \epsilon_{p}\, \varphi_p & = &  - \sum_{\triangle}  \sum_{p,e \in \triangle} \varphi_p   \, \frac{\partial   \alpha^{\triangle}_p}{\partial l_e} \lambda_e  \,+ {\cal O}(\varepsilon^3)
\ea

The kinetic scalar field term is already quadratic, whereas we need to expand $\epsilon_{p}$ to first order in length fluctuations. This can be straightforwardly done and gives the following contribution for a rectangle built out of two triangles  
\ba\label{6.6}
S_{R,\rm{rec}}&=&\frac{1}{AT}\bigg( \varphi_1  \left(  \sqrt{A^2+T^2}\,  \eta_1 - A \alpha_3 -T \tau_2 \right)
+ \varphi_3 \left(  T \tau_1 + A \alpha_3 - \sqrt{A^2+T^2} \,\eta_1\right)\nn\\   &&\q\;
+\varphi_4 \left(  \sqrt{A^2+T^2} \,\eta_1 -A \alpha_1 - T \tau_1    \right)
+\varphi_2 \left(   A \alpha_1 + T \tau_2  -  \sqrt{A^2+T^2} \,\eta_1 \right) \bigg)\q .
\ea
Here $\lambda_p$  with $\lambda=\tau,\alpha,\eta$ denotes the length perturbation associated to an edge $e$ with source vertex $p$, see Figure \ref{fig_Li}. The form of the action (\ref{6.6}) motivates the introduction of 
 the rescaled variables $\hat \lambda_e = \frac{L_e}{\sqrt{6 V_{\sigma}}} \lambda_e$.  
Using the same Fourier transform as in section \ref{Sec:FH} we can write the second order discrete action as  
\ba
S_{\rm L,d} \,=\, \sum_{\rm{rec}} \left(S_{K,\rm{rec}}+ c_{\rm d} S_{R,\rm{rec}}\right)
 \,=\, \frac{1}{2} \sum_{k,\nu}  ((\varphi, \hat \lambda)(k,\nu))^{\rm t} \cdot  \tilde L(k,\nu)  \cdot ((\varphi, \hat \lambda)(-k,-\nu))\ea
with a matrix
\ba
\tilde L(k,\nu)
&=&
 \begin{scriptsize}
 \left(
 \begin{matrix}
  2 \frac{T}{A} \Delta_k + 2 \frac{A}{T} \Delta_v & -c_{\rm d} \Delta_k \frac{(\omega_k^{-1}-\omega_v)}{(1-\omega_k^{-1})}  & - c_{\rm d}(1-\omega^{-1}_v)(-1+ \omega_v\omega_k) & -c_{\rm d}\Delta_k\frac{(-1+\omega_v)}{(1-\omega_k^{-1})} \\
 -c_{\rm d} \Delta_k \frac{(\omega_k-\omega^{-1}_v)}{(1-\omega_k)} & 0&0&0\\
- c_{\rm d}(1-\omega_v)(-1+ \omega^{-1}_v\omega^{-1}_k) & 0 & 0&0\\
 -c_{\rm d}\Delta_k\frac{(-1+\omega^{-1}_v)}{(1-\omega_k)}  &0&0 &0
\end{matrix}
 \right)
 \end{scriptsize}
\ea
where we absorbed into the discrete coupling constant $c_{\rm d}$ the volume factors $\sqrt{6 V_{\sigma}}$.

Note that the matrix entries corresponding to the $\varphi \cdot \hat \lambda_{\rm b}$ coefficients  in the first row (and thus first column) match  --  modulo a $(- c_{\rm d})$ factor -- the $\hat \lambda_r \cdot \hat \lambda_{\rm b}$ entries of the matrix $\tilde M_r$ in (\ref{Hess3}), which we obtained after integrating out  the variables associated to the time axis and the diagonal bulk edges. This confirms the interpretation of the radial bulk edges as a Liouville type field. We have however in $\tilde M_r$ (to the order $x$, which is relevant here) only the Laplacian $\Delta_v$ in time direction appearing. Indeed we will see that we have to modify the Liouville field action (\ref{Lfield1}) by dropping the Laplacian part for the angular direction from the kinetic term of the Liouville field.

 We transform to metric perturbations $\delta h$ as defined in (\ref{htrafo2}).  The action is then encoded in the matrix  
\ba
 \tilde L^{\rm h}(k,\nu)&=&
 \begin{scriptsize}
 \left(
 \begin{matrix}
 2 \frac{T}{A} \Delta_k + 2 \frac{A}{T} \Delta_v & \frac{c'_{\rm d}}{2}  \Delta_k &  \frac{c'_{\rm d}}{2}  \Delta_v & c'_{\rm d} (1-\omega_v)(1-\omega_k)\\
   \frac{c'_{\rm d}}{2}  \Delta_k  & 0&0&0\\
    \frac{c'_{\rm d}}{2}  \Delta_v &0&0&0\\
    c'_{\rm d} (1-\omega^{-1}_v)(1-\omega^{-1}_k) &0&0&0
 \end{matrix}
 \right)
 \end{scriptsize}
\ea
where again we absorb a volume factor $\sqrt{6 V_{\sigma}}$ into a new coupling constant $c'_d$.
Note that the background length parameters $A$ and $T$ only appear in the Laplacian term for the Liouville field $\varphi$. This field can be integrated out and gives rise to an effective action encoded in the matrix
\ba
 \tilde L^{\rm h}_{\rm b}(k,\nu)&=&
- \frac{(c'_{\rm d})^2  }{8 AT (A^{-2} \Delta_k + T^{-2} \Delta_v)}
  \begin{scriptsize}
 \left( 
 \begin{matrix}
 \Delta_k^2 & \Delta_k \Delta_v &  2(1-\omega_k)(1-\omega_v)\Delta_k\\
 \cdots &  \Delta_v^2 &  2(1-\omega_k)(1-\omega_v)\Delta_v\\
 \cdots &\cdots & 4\Delta_k \Delta_v
  \end{matrix}
 \right)
 \end{scriptsize}  \q .
\ea
We see that we match the $x^{-1}$ part (or boundary diffeomorphism invariant part) of the effective boundary action
\ba
\tilde M^{\rm h}_{\rm b}(k,\nu) &=&
-\frac{1}{8x \cdot(6 V_{\sigma})}  {\scriptsize
\left(
\begin{matrix} 
\frac{\Delta^2_k }{\Delta_v}    & \Delta_k   &  2  \frac{ \Delta_k   (1-\omega_k)}{(1-\omega_v^{-1})} \\
\cdots & \Delta_{v} &         2(1-\omega_k)(1-\omega_v) \\
\cdots & \cdots & 4 \Delta_k
\end{matrix}
\right)   
}  + \cdots
\ea
if we omit the angular Laplacian $\Delta_k$ from the kinematic term of the Liouville field action and choose the coupling as
\ba
(c'_d)^2=\frac{A}{T x \, 6 V_{\sigma}} \q .
\ea
Interestingly, by starting from the AdS case and taking a limit to flat spacetimes, \cite{barnichLiou} identified a Liouville field with only time derivatives in the kinetic term as a BMS invariant conformal field theory.

\section{Boundary effective action in the continuum} \label{Sec:Cont}

In this section we will compute the effective boundary action to second order directly in the continuum and confirm the result obtained with the help of the discretization.

To compute the boundary effective action for the linearized theory we have to expand the Einstein--Hilbert action 
\ba\label{8.1}
8 \pi G\,S_{EH}&=&- \frac{1}{2} \int_M d^{d} x \, \sqrt{g}  R \,\,-\,\, \int_{\partial M} d^{d-1} y \sqrt{h} K   \q .
\ea
to second order in variations  $g_{\mu\nu}^{\rm{full}}=g_{\mu\nu} +\delta g_{\mu\nu}$.  The first order variation of the Einstein--Hilbert action is given by (see Appendix \ref{AppFO} for a proof)
\ba\label{8.2}
8 \pi G\,\delta S &=&
-\frac{1}{2}\int_M d^d x \, \sqrt{g}\left( \frac{1}{2}R g^{\mu\nu} -R^{\mu\nu}\right) \delta g_{\mu\nu} 
-\frac{1}{2}  \int_{\partial M} d^{d-1} y \sqrt{h} \left(   K h^{\mu \nu}\  -  K^{\mu \nu}   \right)\delta g_{\mu \nu}  \q .\q\q
\ea
Thus we get for a flat background and a diffeomorphism induced deformation of the metric $\delta g_{\mu\nu}=\nabla_\mu \xi_\nu + \nabla_\nu \xi_\mu$
\ba\label{8.3}
8 \pi G\,\delta S_{\rm flat\, sol}[ \nabla_\mu \xi_\nu + \nabla_\nu \xi_\mu] &=&- \int_{\partial M} d^{d-1} y \sqrt{h} \left(   K h^{\mu \nu}\  -  K^{\mu \nu}   \right)\nabla_\mu \xi_\nu \nn\\
&=&- \int_{\partial M} d^{d-1} y \sqrt{h} \left(   K h^{\mu \nu}\  -  K^{\mu \nu}   \right) ( D_\mu (h_\nu^\rho\xi_\rho)  + K_{\mu\nu} n_\rho \xi^\rho) \q ,
\ea
where we used that
\ba\label{8.4}
h^{\alpha \lambda} h^{\beta\tau} \nabla_\lambda \xi_\tau &=& D^\alpha(h^{\beta \tau} \xi_\tau) + K^{\alpha\beta} n^\tau\xi_\tau \q .
\ea

In our case we have $K^2-K^{\mu\nu}K_{\mu\nu}=0$  as well as  $D^\alpha K^{\mu\nu}=0$ and thus (\ref{8.3})  gives an integral over a  total divergence.   This matches the discrete case \eqref{HJ1}--\eqref{BdryTimeDiff} where the boundary term evaluated on a diffeomorphism induced metric fluctuation vanishes. 
~\\

For the second order variation we do not need to expand the bulk term in (\ref{8.2}), as one can show that the variation of the bulk term vanishes,
 \ba
\delta\left(\sqrt{g}\left( \frac{1}{2}R g^{\mu\nu} -R^{\mu\nu}\right)\right)_{\rm flat\, sol} \,=\, 0   \q .
\ea 
for a flat background and a variation $\delta g_{\mu\nu}=\nabla_\mu \xi_\nu + \nabla_\nu \xi_\mu$.
For the variation of the boundary term in (\ref{8.2}) we find (see Appendix \ref{AppSO})
\ba
&&\delta( \sqrt{h}( K h^{\mu \nu} -K^{\mu\nu}))\nn\\
&=&
\sqrt{h}  \left(
\frac{1}{2}(K h^{\mu \nu} -K^{\mu\nu})g^{\lambda \tau} - K h^{\mu \tau} h^{\nu \lambda}    - h^{\mu\nu} K^{\lambda \tau}  
+ 
2 h^{\lambda(\mu} K^{\nu) \tau} 
\right)\delta g_{\lambda \tau} \nn\\
&&
+\frac{1}{2}\sqrt{h}
\left(  
(  h^{\mu\tau} h^{\nu \lambda}-h^{\mu \nu}  h^{ \lambda \tau} )n^\kappa (\nabla_\tau \delta g_{\lambda \kappa} +\nabla_\lambda \delta g_{\tau \kappa}- \nabla_\kappa \delta g_{ \lambda \tau} ) \right)  \q .
\ea
For a diffeomorphism induced variation of the metric this can be rewritten such that no derivatives $n^\lambda \nabla_\lambda \xi_\tau$ normal to the boundary appear:
\ba\label{8.7}
&&\delta( \sqrt{h}( K h^{\mu \nu} -K^{\mu\nu}))_{\rm diffeo\,\, deform} \nn\\
&=&
\sqrt{h}  \left(
(K h^{\mu \nu} -K^{\mu\nu})h^{\lambda \tau} - 2K h^{\mu \tau} h^{\nu \lambda}    - h^{\mu\nu} K^{\lambda \tau}  
+ 
h^{ \lambda (\mu} K^{\nu) \tau} 
 +  
2  h^{\tau(\mu } K^{\nu)\lambda} 
          \right)\nabla_\lambda \xi_\tau \nn\\
&&
-\sqrt{h}
\left(  
(  h^{\mu\tau} h^{\nu \lambda}-h^{\mu \nu}  h^{ \lambda \tau} )n^\kappa
R_{\kappa(\lambda\tau)\rho}\,\xi^\rho\right)
\nn\\&&
 +\sqrt{h}
\bigg(  
\!\!(D^{(\mu} D^{\nu)} -  h^{\mu \nu} D_\lambda D^\lambda)(n^\rho \xi_\rho) 
-(    K^{\kappa(\mu} D^{\nu)} -  h^{\mu\nu} K^{\kappa\lambda} D_\lambda)(h^\rho_\kappa \xi_\rho)+\nn\\
&&\q\q\q -( h^\rho_\kappa \xi_\rho) ( D^{(\mu} K^{\nu)\kappa}  -  h^{\mu \nu} D_\lambda K^{\lambda\kappa}) 
\bigg)   \q .
\ea
Again we can convert $\nabla_\lambda$ derivatives into $D_\lambda$ derivatives using (\ref{8.4}). In summary we get for the second order of the Hamilton--Jacobi functional evaluated on boundary data induced by an infinitesimal diffeomorphism $\delta g_{\mu\nu} =\nabla_\mu \xi_\nu + \nabla_\nu \xi_\mu$
\ba\label{7.8a}
&&-8\pi G S^{(2)}[ \nabla_\mu \xi_\nu + \nabla_\nu \xi_\mu] \nn\\
&=&
\frac{1}{2}\int_{\partial M} d^{d-1} y 
\sqrt{h}  \Bigg(\left(
(K h^{\mu \nu} -K^{\mu\nu})h^{\lambda \tau} - 2K h^{\mu \tau} h^{\nu \lambda}    - h^{\mu\nu} K^{\lambda \tau}  
+ 
h^{ \lambda (\mu} K^{\nu) \tau} 
 +      
2  h^{\tau(\mu } K^{\nu)\lambda} 
          \right)  \times \nn\\
          &&\q\q\q \q\q\q\q\q\q (  D_\lambda \xi^{||}_\tau + K_{\lambda \tau}  \xi^\perp )   \q\q+\q\q \nn\\
 &&
 \q\q\q \q\q\q\q\q    \left(  (D^{(\mu} D^{\nu)} -  h^{\mu \nu} D_\lambda D^\lambda) \xi^\perp 
-(    K^{\kappa(\mu} D^{\nu)} -  h^{\mu\nu} K^{\kappa\lambda} D_\lambda)\xi_{\kappa}^{||}\right) \Bigg)\times \nn\\
     &&      \q\q\q \q\q\q\q\q  \left( D_{(\mu}   \xi_{\nu)}^{||} + K_{\mu\nu}  \xi^\perp \right)
\ea
where we assumed a flat background metric and vanishing boundary derivatives for the extrinsic curvature tensor. Here we introduced the splitting $\xi^{||}_\tau= h^\rho_\tau \xi_\rho$ and $\xi^\perp = n^\rho \xi_\rho$.


To compare our result to the discrete calculation in section \ref{Sec:Climit} we consider the metric data
\ba \label{MetricData}
g_{\mu\nu}= {\rm{diag}}(1, T^2 , ( A/R)^2 r^{2})  \q,\q\q  h_{\mu\nu}={\rm{diag}}(0, T^2, (A/R)^2 r^{2})   \q .
\ea

We denote the coordinates by $\mu=(r,\tilde t, \tilde \phi)$ where the tilde reminds us of  the rescaling of the standard cylindrical coordinates. The extrinsic curvature is then given as
\ba
K_{\mu\nu}= {\rm{diag}}(0,0 ,  (A/R)^2 r)  \q .
\ea
For the diffeomorphism induced variations we have the following projection onto the boundary
\ba
h_\alpha^{\lambda} h_\beta^{\tau} (\nabla_\lambda \xi_\tau+ \nabla_\tau \xi_\lambda) &=& D_\alpha(h_\beta^{\tau} \xi_\tau) + D_\beta(h_\alpha^{\tau} \xi_\tau)  + 2 K_{\alpha\beta} n^\tau\xi_\tau \nn\\
\ea
that is
\ba\label{7.12}
\delta h_{\mu\nu} &=&
\left(
\begin{matrix}
0&0&0\\
0&2\partial_{\tilde t} \xi_{\tilde t}  & \partial_{\tilde t} \xi_{\tilde \phi} +  \partial_{\tilde \phi} \xi_{\tilde t} \\
0&  \partial_{\tilde t} \xi_{\tilde \phi} +  \partial_{\tilde \phi} \xi_{\tilde t} \,\,\,& 2\partial_{\tilde \phi} \xi_{\tilde \phi} + 2 \tilde A^2 r \xi_r
\end{matrix}
\right) \q .
\ea
We will later translate the derivatives $\partial_{\tilde t}$ and $\partial_{\tilde \phi}$ to the wave vectors $\hat k$ and $\hat  v$ and indeed see that these metric perturbations induced by  diffeomorphisms in the various directions, described by (\ref{7.12})  do match the continuum limit we found for the metric perturbations in the discrete case, see (\ref{diffeoB}) and (\ref{diffeoR}).

Evaluating the second order action (\ref{7.8a}) for such metric perturbations as in (\ref{7.12}) we obtain
\ba\label{7.13}
8\pi G \,S^{(2)}[ \nabla_\mu \xi_\nu + \nabla_\nu \xi_\mu] 
&=&-\frac{1}{2} \int d\tilde t d\tilde \phi
 \left( - \frac{ A}{RT}\xi_r \partial^2_{\tilde t} \xi_r +   \frac{ A}{R T^3}\xi_{\tilde t} \partial^2_{\tilde t} \xi_{\tilde t} + \frac{R}{ A  T r^2}\xi_{\tilde \phi} \partial^2_{\tilde t} \xi_{\tilde \phi}\right) \q\q 
\ea
where we used integration by parts, that lead to the cancellation of many terms.

 To finally compare to the discrete calculation in section \ref{Sec:Climit} we define the Fourier transform
\ba\label{7.14}
f(k,\nu) &=&  \sqrt{\frac{ A T}{2\pi R \beta} }\int_0^{2\pi R/ A} \!\! d\tilde \phi  \int_0^{\beta/ T}\!\!  d\tilde t \,\, \exp\left(-i   \left( \frac{A}{R} \tilde \phi k + \frac{2\pi}{\beta } T \tilde t (  \nu-\frac{\gamma}{2\pi}k) \right) \right)\,\, f(\tilde t, \tilde \phi) \q .\q\q
\ea
This again incorporates the periodicity $(\tilde t, \tilde \phi) \sim (\tilde t + \beta/ T, \tilde \phi + \gamma R / A)$ of the functions on the twisted torus via the introduction of $v:=( \nu-\frac{\gamma}{2\pi}k)$.  The inverse Fourier transformation is given as
\ba
f(\tilde t, \tilde \phi) \,=\,\sqrt{ \frac{  A  T}{2\pi R \beta}} \sum_{k,\nu \in \mathbb{Z}}  \exp\left(i   \left(   \frac{A}{R} \tilde \phi k + \frac{2\pi}{\beta } T \tilde t (  \nu-\frac{\gamma}{2\pi}k) \right) \right) \,\, f(k,\nu) \q .
\ea

Defining $\hat v=\frac{2\pi}{\beta}  T v$ and $\hat k=(A/R) k$ we  rewrite (\ref{7.13}) into
\ba
8\pi G \,S^{(2)}[ \nabla_\mu \xi_\nu + \nabla_\nu \xi_\mu] 
&=&
\frac{1}{2} \sum_{k,\nu \in \mathbb{Z}}  \left( -\frac{ A}{R T} \hat v^2  |\xi_r|^2 +   \frac{ A}{R T^3} \hat v^2  |\xi_{\tilde t}|^2 + \frac{R}{ A  T r^2} \hat v^2 |\xi_{\tilde \phi}|^2 \right)  \q ,
\ea
which indeed matches the continuum limits (\ref{5.43}) and (\ref{5.46}) of the discrete calculation.

Thus we confirm the calculation of the effective boundary action in the discrete case. The calculation here is easily generalizable to other boundaries and one can again ask for a dual field theory. This  will be subject for future work.

\section{Outlook}\label{outlook}

Here we point out some possibilities for future work. One issue is the recourse to an alternative discretization, necessary if one adds (non--topological) matter fields. Using such an alternative discretization gives also a connection to MERA tensor networks and holographic renormalization. Another issue is the implementation of the alternative boundary term $\frac12 B_{GHY}$ used in \cite{Barnich1}. To this end  one also has to adapt the discretization. 

\subsection{Alternative  discretizations: refinement by radial evolution}

In this work we computed the partition function for 3D Regge calculus at one--loop, making use of the topological nature of the theory. Thus we used a very coarse bulk triangulation, which minimized the computational effort. Despite the topological nature of the theory we found that the partition function with boundary can be described by a field theory. In particular the part of the effective boundary action that is invariant under (linearized) boundary diffeomorphisms is dual to a field theory of Liouville type. 

The topological invariance of the 3D Regge action and the measure (\ref{Rmeasure}) ensures that the results do not change if we alter the bulk triangulation. In particular we could change our set--up  and subdivide the cylinder, obtained by cutting the torus, into cylindrical rings and a thin cylinder at the centre. For the cylinder at the centre we would use the same triangulation as before and of course obtain the same partition function. The cylindrical rings are again cut into time slices, and these slices are subdivided by alternating cuboids and the same prisms as considered before. (The Hessian of the Regge action for such cuboids has been computed in \cite{dfs}.)

\begin{figure}
\includegraphics[scale=0.5]{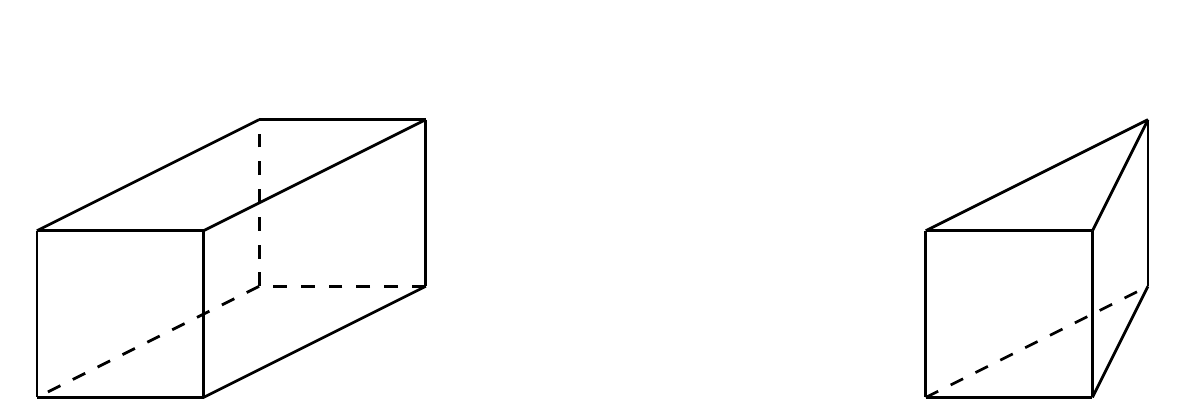} \q\q\q\q\q\q\q\q\q\q
\includegraphics[scale=0.5]{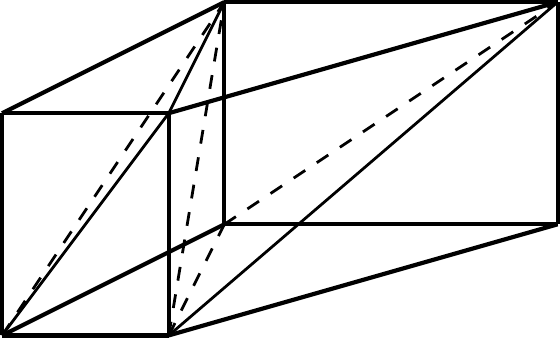}
\caption{\small  A cylindrical ring can be discretized using prisms and cuboids in alternating order. This leads to a refinement of the outer boundary. Alternatively one can use cuboids with more general choice of background lengths. These are also needed if one accommodates non--vanishing curvature induced by an alternative boundary term.}\label{cube}
\end{figure}

The partition function for a cylindrical ring does now depend on two sets of boundary data, from the inner and outer boundary. We can thus understand this partition function as a map from the Hilbert space defined on the inner boundary to a Hilbert space defined on the outer boundary.   In the set--up considered here the outer boundary would have more lattice sites. Note that it would not be characterized by a ``small radius'', as the radius is part of the boundary variables over which one would integrate in glueing the partition function.  The map defined by the cylindrical ring realizes therefore a refinement map, as discussed in \cite{bd12,bdsteinhaus13,bd14}. Moreover this refinement map is defined by the dynamics itself -- one uses the ``radial evolution'' map for the definition of this refinement map. 

Also  MERA tensor networks \cite{MERA} can be essentially seen as refinement maps between Hilbert spaces of different sizes. Moreover the two building blocks of such a MERA networks, disentanglers and isometries, are mirrored in the cuboids and prisms, with which one can discretize the cylindrical rings. MERA networks have been discussed lately also in the relation to  AdS/CFT holography \cite{AdSMERA}. In particular it has been proposed that the entanglement structure encoded in the networks defines an AdS geometry. See however \cite{AltMERA} for alternative proposals. The Regge discretization discussed here can also be understood in terms of tensor networks, in particular if one generalizes to so--called decorated tensor networks introduced in \cite{dectnw}.  This tensor network would a priori be a 3D network, as opposed to the MERA networks, which are two--dimensional.  In fact the MERA networks are proposed as describing stationary spacetimes. We can however produce a 2D network from the 3D network by considering only one time slice and by gluing the down and up sides of each building block. In this we can produce an example (that provides the tensors explicitly) of a MERA network describing gravity. Here we are concerned with the case without cosmological constant, however this is easily generalizable to gravity with a (negative) cosmological  constant.

The discretization of the torus using cuboids and prisms further allows to perform the continuum limit in the bulk. This will not change the results for a topological theory, as considered here, however it will be necessary if one considers non--topological theories, for instance by adding matter field to the gravitational dynamics.  Using a negative cosmological constant the set--up is  related to that of holographic renormalization \cite{HolR} and the maps defined by the cylindrical rings can be understood to implement the renormalization flow.

\subsection{Alternative  boundary term}

To connect the partition function for the torus to thermodynamical quantities, it is necessary to choose a different boundary term which is 1/2 of the usual GHY boundary term given in \eqref{GHY}, \cite{detournay}. In particular, \cite{detournay} specifies asymptotic boundary conditions at infinity which lead to a well--defined variational principle.  Regge calculus, on the other hand, incorporates naturally the GHY boundary term without the extra $1/2$ \cite{sorkinBdry}. One can adjust the Regge action by adding terms so that in effect one has 1/2 of the GHY term. This does however change the equations of motion for edges which are contained in tetrahedra glued to the boundary.

For the triangulation at hand this does however apply for all edges.  Indeed  the background equations  of motion get changed for the edges making up the  time axis and the radial edges, which are now required to have non--vanishing deficit angle. To avoid this we can change to a triangulation built in the following way: we take our original triangulation (with only prisms) and glue to it a `thin' ring. This ring  can be either triangulated with only cuboids or with cuboids and prisms, as we described in the previous section. 

The equations of motion will now require non--vanishing deficit angles around the time and radial edges of the cylindrical ring. To accommodate them (that is in particular a non--vanishing deficit angle for the radial edges) we need to allow more general background lengths for the cuboid (and prism) building blocks, see Figure \ref{cube}.

In this way we can use the cylindrical ring discretization as a map which convert the standard boundary term into the alternative boundary term. After computing the partition function at finite radius one can then consider the infinite radius limit.

Alternatively one can ask for boundary conditions at finite radius which are compatible with the alternative boundary term. In general this will involve  keeping a combination of the boundary metric and the extrinsic curvature fixed. This can be also incorporated into Regge calculus, either by working with length variables, or, by changing to a first order version of the theory \cite{barret, newangle}.  In the first order version one uses both the dihedral angles and the length variables, which translate to extrinsic curvature and metric variables in the continuum.  (Another version \cite{areaangle} of (4D) Regge calculus uses area and angle variables which match the semiclassical variables  of  (4D) spin foams.) 

In all those cases one can work with the alternative variables for the full triangulation or use standard length variables and then transform the resulting partition function to new boundary conditions. This transformation can again be implemented by gluing a cylindrical ring  to the triangulation, where for the inner boundary one uses the standard boundary term and length variables and for the outer boundary the alternative boundary term and boundary conditions.

\section{Discussion}\label{discussion}

In this work we aimed at an exploration of holographic dualities in non--perturbative approaches. For that, it is important to generalize the considerations from boundaries at asymptotic infinity to boundaries of finite size. As an example we considered 3D flat gravity, as the corresponding spin foam model, the Ponzano--Regge model \cite{PR},  is the one most  understood and under control, in particular compared to the AdS case. 

We computed the one--loop partition function using (quantum) Regge calculus, which constitutes the semi--classical limit of the Ponzano--Regge model. We showed that it is indeed possible to work at finite boundaries. The evaluation of the classical action is completely independent of the (background) radius at which the boundary is situated. The part of the one--loop determinant which depends non--trivially on the twisting angle $\gamma$ also arises for arbitrarily small radius: This part can be understood to arise from a ``dual'' scalar field with a Lagrangian kinetic term $\phi \partial^2_t \phi$. This ``dual'' field can be identified with the degrees of freedom attached to the radial edges.  The $\gamma$--dependent part of the one--loop determinant is given by the determinant of $\partial_t^2$, which acquires a non--trivial value due to the twisting angle. This is a topological feature, independent of the value of the radius.

Let us point out that the use of a discretization, and in particular the geometric features of Regge calculus, allowed an immediate interpretation of the structure of the one--loop result, as compared to the continuum. The discretization makes use of the fact that one deals with a topological theory and makes the computation of the one--loop determinant possible by simple algebraic means.

In general we wish to emphasize that the full dynamics of quantum gravity is encoded in its partition function as a functional of (arbitrary) boundary data.  For approaches where quantum gravity is defined holographically via a boundary theory, the question arises whether this can be extended to ``non--asymptotic'' boundaries. For discrete approaches, the challenge is to allow a continuum limit for the boundary data. We have seen here that this can lead to non--trivial results even in the case of topological theories.

There are numerous further directions and open questions left for future work:
\begin{itemize}
\item An immediate question is to consider the partition function as given by the Ponzano--Regge model.  Work in this direction is in progress \cite{AEtoappear}. Regge calculus arises as the semi--classical limit of the Ponzano--Regge model. Diffeomorphism symmetry is furthermore implemented in the same way as in Regge calculus \cite{louapre}. Therefore one would expect a similar result. There are however important open questions: (a) The semi--classical limit involves large spins, which means large lengths. The question is whether this interferes with the continuum limit at the boundary. Therefore this model could constitute an interesting case study for the continuum limit in spin foams and loop quantum gravity \cite{eckert,bd14} in general. Note that the partition function for a solid torus has been considered in \cite{Hellmann}, however in a very coarse discretization that does not allow to capture the twisting angle. (b) The Ponzano--Regge model is a fully quantum object which works with {\it a priori} complex weights $\exp(iS)$. It also incorporates the sum over orientations, which in effects leads to real weights $\exp( iS)+\exp(-iS)$. This feature circumvents the conformal factor problem. An important question is however if the sum over orientations changes the partition function (even in a semi--classical limit) as compared to working with metric variables or excluding orientation changes \cite{antispacetime}.

Concerning possible dual field theories, the Ponzano--Regge model is indeed related to the Ising model defined on the boundary triangulation \cite{BonzomLivineIsing}, see also \cite{BDHnybida}. Another approach to 3D gravity is BF theory \cite{Witten3D}, of which the Ponzano--Regge model is a direct quantization. The corresponding partition function is known to be an integral of the Ray--Singer torsion over the moduli space of flat connections \cite{TorsionBF} (or of the Reidemeister, or combnatorial, torsion at the discrete level \cite{TorsionPR}). For the case considered here one needs to include a boundary and keep the $B$--field on the boundary fixed.

\item BTZ black holes \cite{btz} lead also to partition functions defined on a solid torus. It has been suggested that the corresponding partition function can be obtained from the Turaev--Viro model, which describes 3D gravity with positive cosmological constant, by an analytical continuation \cite{marc}. Here we pointed out the importance of considering the continuum limit on the boundary, in particular if one wishes to obtain a description in terms of a dual field. This dual field description is suggested to be responsible for the black hole entropy \cite{carlip}. 

\item We considered the case of 3D gravity without cosmological constant. An important feature we used is that 3D (standard) Regge calculus is discretization--independent in the bulk. This can be extended to the measure term, at least for the linearized theory \cite{Meas}.  Adding a cosmological constant term does break the triangulation independence as well as diffeomorphism invariance of standard 3D Regge calculus \cite{dittrich08}. This can however be avoided by changing to a version of Regge calculus, in which the flat building blocks are exchanged for homogeneously curved building blocks \cite{improved,newangle}. This restores triangulation independence and diffeomorphism invariance and also the path integral measure  can be adjusted to the curved case.

\item In this work we computed the Hamilton--Jacobi functional for linearized gravity for a torus (finite) boundary. Given that in non--perturbative approaches one is interested in the partition function for arbitrary boundaries, it is worthwhile to generalize this study of the Hamilton--Jacobi functional to other boundaries and to also include a cosmological constant \cite{ABS}. One question is whether one can also in those cases extract a dual field theory on the boundary.

\item Here we considered a topological theory, which allowed us to employ the coarsest possible bulk triangulation, while taking the continuum limit only on the boundary. Adding matter fields or considering 4D gravity would change this topological nature. A continuum limit for the bulk triangulation would then be required as well. To do so for 3D gravity, we have to change the triangulation, for instance by using cylindrical rings, as described in section \ref{outlook}. It should be still straightforward to evaluate the corresponding partition function for linearized Regge calculus. Allowing (matter) excitations in the bulk one can study the bulk--boundary correspondence within a concrete model. Adding point particles as matter would still keep the topological nature of the theory and a corresponding bulk--boundary correspondence has been suggested in \cite{krasnovPP}.

\item  The gravity partition function with boundary solves the Wheeler--DeWitt equation. (On the classical level the Hamilton--Jacobi functional solves the Hamilton--Jacobi functional equation.) On the other hand we can use the Wheeler--DeWitt equation to characterize the partition function and derive its properties. The Wheeler--DeWitt equation can come in different forms, as a functional differential equation in terms of the boundary metric or as a difference equation in the spin values for the Ponzano--Regge model. The Wheeler--DeWitt equation for the Ponzano--Regge model is provided in \cite{BarrettCrane, BonzomFreidel}  for  three--valent vertices, for the example in this work we need it for a  lattice with six--valent vertices and possibly expanded on background values corresponding to a torus geometry. The Wheeler--DeWitt equation can be also implemented by studying how the partition function changes under a change of the boundary triangulation \cite{hoehn2}. In particular  one can consider a sequence of triangulation changes that brings one back to the original triangulation, so--called tent moves \cite{tent,hoehn2}. Tent moves define an evolution normal to the boundary and in this sense are  again related to the  Wheeler--DeWitt equation. Similarly a gluing of a cylindrical ring to the triangulation can be translated to  a (global) ``radial'' evolution and a radial Hamilton--Jacobi equation, which gives the holographic renormalization flow \cite{HolR}. Here one can consider a gluing of the cylindrical ring that either refines the boundary triangulation (using prisms and cubes) or keeps the same boundary triangulation (using only cuboids).

\item A topic we did not touch in this work are the BMS symmetries at asymptotic infinity. An interesting question is to consider the case of finite boundaries. For the AdS case, a first study appeared recently \cite{Marolf}. In addition we are using here a discretization, and the question arises whether and how this is compatible with the BMS symmetry (before implementing a continuum limit). That a symmetry algebra can still be recovered for discrete geometries has been shown for the 3D hypersurface deformation algebra in \cite{bonzomdittrich}.

\end{itemize}

\begin{acknowledgments}
We are very thankful to  Laurent Freidel, Ted Jacobson, Etera Livine, Rob Myers and  in particular Aldo Riello  for discussions. We thank Perimeter Institute for hosting the workshop "Renormalization in background independent theories" and its participants for enlightening remarks.
This research was supported in part by Perimeter Institute for Theoretical Physics. Research at Perimeter Institute is supported by the Government of Canada through Industry Canada and by the Province of Ontario through the Ministry of Research and Innovation.
\end{acknowledgments}

\begin{appendix}
\section{The Hessian for the Regge action} \label{sec:appendixA}

For the evaluation of the Hessian for the Regge action we need the length derivatives of the dihedral angles. A general formula for the dihedral angles in a $d$--dimensional simplex has been derived in \cite{dfs}:
\ba\label{magic1}
\frac{\partial \theta(kl)}{\partial l_{hm}} &=&\frac{1}{d^2} \frac{l_{hm}}{\sin \theta_{kl}} \frac{V(h) V(m)}{V^2} 
\big(
(\cos\theta(kh)\cos\theta(ml)+\cos\theta(km)\cos\theta(hl))
+
 \nn \\
&& \q\q\q\q\q\q\q\q
\cos\theta(kl)\,(\cos\theta(kh)\cos\theta(km)+\cos\theta(lh)\cos\theta(lm))
\big) 
\ea
To explain the notation we label the vertices of the $d$--dimensional simplex with $k=1,\ldots, d+1$. The (interior) dihedral angle $\theta(kl)$ is the angle between the faces of the simplex, obtained by removing the vertices $k$ and $l$ respectively. For $d=3$ the angle $\theta(kl)$ is hinging at the opposite edge to the one between the vertices $k$ and $l$. The length $l_{hm}$ is the length of the edge between $h$ and $m$, $V$ is the volume of the $d$--simplex and  $V(h)$ is the volume of the $(d-1)$--simplex obtained by removing the vertex $h$. 

By convention we have $\cos\theta(ii)=-1$, which leads to simplifications of (\ref{magic1}) in case  that one considers a derivative of a dihedral angle with respect to the length of the opposite edge, i.e.
\ba
\frac{\partial \theta(kl)}{\partial l_{kl}} &=&\frac{l_{kl}}{d(d-1)}\frac{V(kl)}{V}
\ea
or the derivative of a dihedral angle with respect to the length of an  edge opposite to an adjacent edge
\ba
\frac{\partial \theta(kl)}{\partial l_{km}} &=&-\frac{l_{km}}{d(d-1)}\frac{V(kl)}{V}\frac{V(m)}{V(l)} \cos\theta(ml) \q .
\ea

\begin{figure}
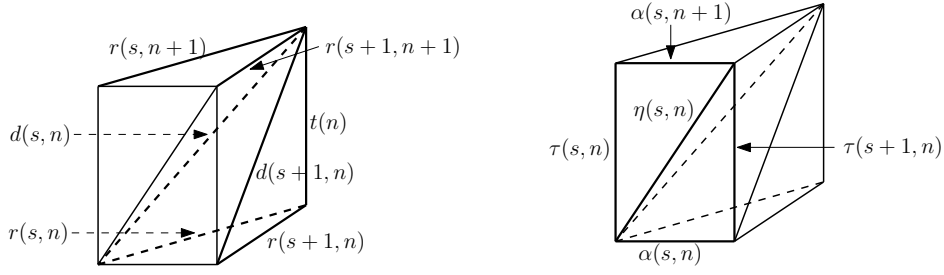

\includegraphics[scale=0.7]{figures/PrismVariables} \q\q\q
\includegraphics[scale=0.7]{figures/PrismBdryVariables.pdf}
\caption{\small The edge variables for the prism.}\label{fig_1b}
\end{figure}

With the help of these formulas one can compute the Hessian of the Regge action associated to the prism building block in Figure \ref{fig_1b}:
\ba
H^{\rm pr}_{ee'} &=& \sum_{\sigma \in {\rm pr}} \sum_{\sigma \supset e,e'}  \frac{\partial  \theta^\sigma_{e} }{\partial l_e} \,\,=\,\,\frac{L_e L_{e'}}{6 V_\sigma}  M^{\rm pr}_{ee'} (x) \q .
\ea
The matrix $ M^{\rm pr}$ is given as
\ba
M^{\rm pr}
&=&
{\scriptsize
\left(
\begin{array}{cccccccccccc}
1-x&-1& 0 & 0 & -1+x & 1 & -x & 0 & 0 & 0 & 0& 0\\
-1&1-x&0&0&1&-1+x&-x& 0& 0& 0& 0& 0\\
0&0&x&0&-x&0&0&-1+x&0&0&-1&1\\
0&0&0&x&0&-x&0&1&x&1&0&-1\\
-1+x&1&-x&0&1&-1&0&0&1&0&1&-1\\
1&-1+x&0&-x&-1&1&0&0&-1&-1&0&1\\
-x&-x&0&0&0&0&0&0&0&1&0&0\\
0&0&-1+x&1&0&0&0&0&0&0&-\tfrac{1}{2}&0\\
0&0&0&x&1&-1&0&0&1&\tfrac{1}{2}&0&-1\\
0&0&0&1&0&-1&1&0&\tfrac{1}{2}&\tfrac{1}{2}&0&-\tfrac{1}{2}\\
0&0&-1&0&1&0&0&-\tfrac{1}{2}&0&0&\tfrac{1}{2}&-\tfrac{1}{2}\\
0&0&1&-1&-1&1&0&0&-1&-\tfrac{1}{2}&-\tfrac{1}{2}&1
\end{array}
\right)
} \q 
\ea
with $x=A^2/(2R^2)$.
Here the entries of the matrix are given in the order 
\ba
&&r(s,n),r(s+1,n),r(s,n+1),r(s+1,n+1),d(s,n),d(s+1,n),t(n),\nn\\&&
\tau(s,n)\tau(s+1,n),\alpha(s,n),\alpha(s,n+1),\eta(s,n)
\ea
with the edge labeling displayed in Figure \ref{fig_1b}.

\section{Length changes induced by vertex displacements}\label{AppDiff}

\subsection{Boundary diffeomorphisms}

\begin{figure}[htb]
\includegraphics[scale=0.5]{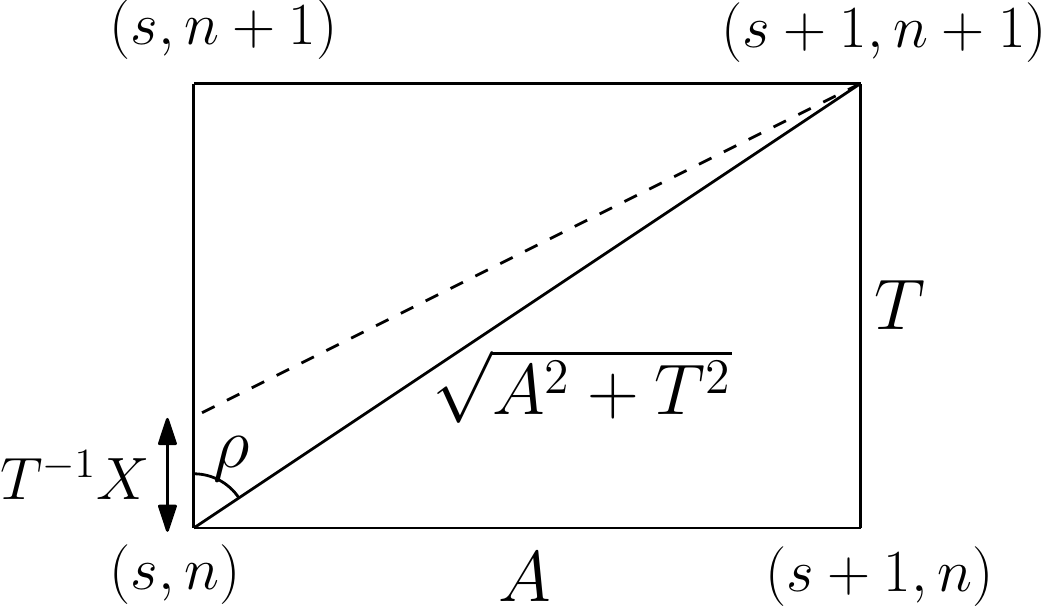}
\caption{\small  A boundary vertex translation in time direction.}\label{fig_Ti}
\end{figure}

Let us consider the boundary geometry only and assume we translate a certain vertex $(s,n)$ by a length amount $T^{-1}X_\tau$ in  the positive time direction. (In this way $X_\tau$ will correspond  to the one--form (component) describing a diffeomorphism based on a vector with length $\sqrt{h^{\tau\tau}}X_\tau$.) This will affect the length of the adjacent edges. To first order in $X_\tau$ edges with an orthogonal angle to the time direction will however not change. The induced change for the edges in time direction is
\ba
\tau(s,n)\,=\, -T^{-1}X_\tau \, , \q \tau(s,n-1) \,=\, T^{-1}X_\tau \q . 
\ea
Furthermore the diagonals at vertices $(s,n)$ and $(s-1,n-1)$ are affected and we have $\eta(s,n)=-\eta(s-1,n-1)$.  The length change of the diagonal at $(s,n)$ can be found using the trigonometric relation according to the situation in Figure \ref{fig_Ti}
\ba
\Bigl(\sqrt{A^2+T^2} + \eta(s,n) \Bigr)^2 &=& A^2+T^2+T^{-2}X^2 - 2 X T^{-1}\sqrt{A^2+T^2} \cos \rho  \q \Rightarrow \nn\\
   \eta(s,n) &=& -X T^{-1}\cos \rho + {\cal O}(X^2)+ {\cal O}(\eta^2)
\ea
Here $\rho$ is the angle between the diagonal and the time edge for which we have $\cos\rho=\frac{T}{\sqrt{A^2+T^2}}$.  Changing to the variables $\hat \lambda_e= \frac{L_e}{\sqrt{6 V_\sigma}}$ and Fourier transforming we find the deformation vector 
\ba
(n_{{\rm b},\tau})^{\rm t}(k,\nu) &=& -\frac{1}{ \sqrt{6 V_{\sigma} } }(1-\omega_v , 0, 1-\omega_v \omega_k) \, X_\tau(k,\nu)  \q .
\ea
In metric perturbation variables $\delta h$ we find
\ba\label{diffT1}
(n^{\rm h}_{{\rm b},\tau})^{\rm t}(k,\nu) &=&-\,\, (2(1-\omega_v) ,\,\, 0,\,\,  \omega_v(1-\omega_k) )  \,  \,\,X_\tau(k,\nu)  \q .
\ea
Likewise we find for a vertex displacement in the angular direction
\ba\label{diffA1}
(n^{\rm h}_{{\rm b},\alpha})^{\rm t}(k,\nu) &=&-\,\, (0 ,\,\, 2(1-\omega_k),\,\,  \omega_k(1-\omega_v) )  \,  \,\,X_\alpha(k,\nu)  \q .
\ea

\subsection{A displacement orthogonal to the boundary}

Consider a  displacement of  a vertex $(s,n)$ on the boundary in radial (outward) direction by some amount $X_r$. We  embed the piecewise linear surface into 3D flat space in order to discuss the induced length change of the boundary edges. This displacement will affect the edges in angular direction at $(s,n)$ and $(s-1,n)$ equally as well as the diagonal edges at $(s,n)$ and $(s-1,n-1)$. 
The induced length change is given to first order by
\ba
\alpha(s,n)&=&\alpha(s-1,n)\,=\,  X_r \sin\Bigl( \frac{\pi}{N_A}\Bigr)  \q, \nn\\
\eta(s,n)&=&\eta(s-1,n-1)\,=\, X_r    \sin\Bigl( \frac{\pi}{N_A}\Bigr) \frac{A}{\sqrt{A^2+T^2}} \q ,
\ea
so that in variables $\hat \lambda$,
\ba
(n_{{\rm b},r })^{\rm t}(k,\nu) &=& \frac{A \sin( \tfrac{\pi}{N_A})}{ \sqrt{6 V_{\sigma} } }(0 , 1+\omega_k, 1+\omega_v \omega_k) \, X_r(k,\nu)  \q .
\ea
In variables $\delta h$ we have
\ba
(n^{\rm h}_{{\rm b},r })^{\rm t}(k,\nu) &=& A \sin( \tfrac{\pi}{N_A})\, (0 ,2( 1+\omega_k), (1-\omega_v^{-1} )\omega_v\omega_k) \, X_r(k,\nu)  \q .
\ea

\section{Variation of the Einstein--Hilbert action with Gibbons--Hawking--York boundary term}
\subsection{First order}\label{AppFO}

We consider the Einstein--Hilbert action with Gibbons--Hawking--York boundary term 
\ba\label{actionC1}
-8\pi G S_{EH}&=& \frac{1}{2} \int_M d^{d} x \, \sqrt{g}  R \,\,+\,\, \int_{\partial M} d^{d-1} y \sqrt{h} K   \q .
\ea
We will first show that the first order variation of this action is given by  (see e.g. \cite{detournay})
\ba\label{C2}
-8\pi G \delta S_{EH} &=&
\frac{1}{2}\int_M d^d x \, \sqrt{g}\left( \frac{1}{2}R g^{\mu\nu} -R^{\mu\nu}\right) \delta g_{\mu\nu} 
+\frac{1}{2}  \int_{\partial M} d^{d-1} y \sqrt{h} \left(   K h^{\mu \nu}\  -  K^{\mu \nu}   \right)\delta g_{\mu \nu}  \, .\q\q
\ea

The variation of the bulk term uses that
\ba
g^{\mu\nu} \delta R_{\mu\nu} \,=\, \nabla^\mu (\nabla^\nu \delta g_{\mu \nu} - g^{\nu \rho} \nabla_\mu \delta g_{\nu \rho}) 
\ea
is a total divergence. This follows from the Palatini equations (which can be proven using Riemann normal coordinates)
\ba\label{Palatinieq}
\delta {R_{\mu\nu\rho}}^\sigma&=&\nabla_\nu\delta\Gamma^\sigma_{\mu\rho}-\nabla_\mu\delta\Gamma^\sigma_{\nu\rho} \, , \nn\\
\delta {R_{\mu\rho}}&=&\nabla_\sigma\delta\Gamma^\sigma_{\mu\rho}-\nabla_\mu\delta\Gamma^\sigma_{\sigma\rho} \, , \nn\\
\delta\Gamma^\mu_{\nu\rho}&=&\frac{1}{2} g^{\mu\sigma}\left(\nabla_\nu \delta g_{\rho\sigma} +  \nabla_\rho \delta g_{\nu\sigma} - \nabla_\sigma \delta g_{\nu\rho}\right) \q .
\ea
Together with the terms resulting from the variation of the inverse metric and the determinant we obtain
\ba
\delta( \sqrt{g} R) &=& \sqrt{g}\left( \frac{1}{2} g^{\mu\nu} R-R^{\mu\nu}\right) \delta g_{\mu\nu}  \,+\, \sqrt{g} \nabla^\mu\left( \nabla^\nu \delta g_{\mu\nu} - g^{\nu \rho} \nabla_\mu \delta g_{\nu \rho} \right) \q .
\ea
For the variation of the boundary term we will need the variation of the normal to the boundary $n$. From
\ba
0\,=\,\delta( n_\mu n_\nu g^{\mu\nu}) &=& 2 n^\mu \delta n_\mu - n^\mu n^\nu \delta g_{\mu \nu}
\ea
we can deduce
\ba
n^\mu \delta n_\mu &=& \frac{1}{2} n^\mu n^\nu \delta g_{\mu \nu} \q .
\ea
This determines the variation of the normal parallel to the normal itself. The variation of the normal orthogonal to the normal itself does however vanish. This follows from the fact that we consider a fixed embedded hypersurface $x^\mu(y^a)$ with tangential vectors $X^\mu_a:=\partial x^\mu/\partial y^a$. Hence
\ba
0\,=\, \delta( n_\mu X^\mu_a) \q \Rightarrow \q X^\mu_a \delta n_\mu \,=\,0  \q .
\ea
Thus, using $ n_\mu n^\rho = g^{\rho}_\mu-h^{\rho}_\mu$ we have
\ba\label{C8}
 \delta n_\mu &=& \frac{1}{2}  n_\mu  n^\rho n^\nu \delta g_{\rho \nu}  \nn\\
 &=&  \frac{1}{2} n^\nu \delta g_{\mu \nu} -\frac{1}{2} h^{\rho}_\mu n^\nu \delta g_{\rho \nu} 
 \,\, =: \,\,  \frac{1}{2} n^\nu \delta g_{\mu \nu} -\frac{1}{2} w_\mu \q ,
 \ea
where for later reference we note that the co-vector $w_\mu$ is orthogonal to $n^\mu$. 

With the definition 
\ba
K_{\mu\nu} \,=\, h_{\mu}^\rho h^\sigma_\nu \nabla_\rho n_\sigma \,=\, h_{\mu}^\rho  \nabla_\rho n_\nu
\ea
we have for the variation of the extrinsic curvature trace
\ba\label{C17a}
\delta K &=& \delta( h^{\mu \nu}  \nabla_\mu n_\nu) \nn\\
&=&  \delta( g^{\mu \nu}  \nabla_\mu n_\nu) \nn\\
&=& - (\nabla_\mu n^\sigma)\, (h^{\rho \mu} +n^\rho n^\mu)  \delta g_{\rho\sigma} \,+\,  g^{\mu \nu} \delta( \nabla_\mu n_\nu) \nn\\
&=& - K^{\mu \nu} \delta g_{\mu \nu} - n^\rho n^\mu (\nabla_\mu n^\sigma) \delta g_{\rho\sigma} \,+\,  g^{\mu \nu} \delta( \nabla_\mu n_\nu)  \q .
\ea
To evaluate $\delta( \nabla_\mu n_\nu)$ we write  $\nabla_\mu n_\nu=\partial_\mu n_\nu- \Gamma^\rho_{\mu\nu}n_\rho$ and use the Palatini equations (\ref{Palatinieq}). 

After several rewritings and employing the boundary covariant derivative $D_\mu v_\alpha=h^\rho_\mu h^\sigma_\alpha \nabla_\rho v_\sigma$ (for a one--form $v_\sigma$ with $n^\sigma v_\sigma =0$) we arrive at
\ba
\delta( \sqrt{h} K) &=&   \frac{1}{2}  \sqrt{h} \left( K h^{\mu \nu}\delta g_{\mu \nu}   -  K^{\mu \nu} \delta g_{\mu \nu} - g^{\mu\nu} n^\kappa \left(  \nabla_\nu \delta g_{\mu \kappa}  -\nabla_\kappa \delta g_{\mu\nu} \right)
-  D^\mu w_\mu \right)  \q .\q\q
\ea

In summary, converting the bulk total divergence to an integral over the boundary and dropping the divergence term $D^\mu w_\mu$, we obtain for the first variation of the action (\ref{actionC1}) 
\ba\label{C17b}
-8\pi G\,\delta S &=&
\frac{1}{2}\int_M d^d x \, \sqrt{g}\left( \frac{1}{2}R g^{\mu\nu} -R^{\mu\nu}\right) \delta g_{\mu\nu} 
+\frac{1}{2}  \int_{\partial M} d^{d-1} y \sqrt{h} \left(   K h^{\mu \nu}\  -  K^{\mu \nu}   \right)\delta g_{\mu \nu}  \q .\q\q
\ea

\subsection{Second order variation of the boundary term}\label{AppSO}

We need to know the action to second order in perturbations and hence need to take a second variational derivative of (\ref{C17b}). As mentioned in the main text one can show that the variation of the bulk term
\ba
\delta \left( \sqrt{g} \left( \frac{1}{2} g^{\mu\nu} R-R^{\mu\nu} \right) \right) 
 \ea
vanishes if evaluated on a flat background and for a diffeomorphism induced perturbation $\delta g_{\mu\nu} =\nabla_\mu \xi_\nu + \nabla_\nu \xi_\mu$.

We are thus left with the variation of the boundary term. In the following we will show that
\ba
&&\delta( \sqrt{h}( K h^{\mu \nu} -K^{\mu\nu}))\nn\\
&=&
\sqrt{h}  \left(
\frac{1}{2}(K h^{\mu \nu} -K^{\mu\nu})g^{\lambda \tau} - K h^{\mu \tau} h^{\nu \lambda}    - h^{\mu\nu} K^{\lambda \tau}  
+ h^{\mu \lambda} K^{\nu \tau} + h^{\nu \lambda} K^{\mu \tau} \right)\delta g_{\lambda \tau} \nn\\
&&
+\frac{1}{2}\sqrt{h}
\left(  
(  h^{\mu\tau} h^{\nu \lambda}-h^{\mu \nu}  h^{ \lambda \tau} )n^\kappa (\nabla_\tau \delta g_{\lambda \kappa} +\nabla_\lambda \delta g_{\tau \kappa}- \nabla_\kappa \delta g_{ \lambda \tau} ) \right)  \q .
\ea

To start, we have
\ba
\delta( \sqrt{h}( K h^{\mu \nu} -K^{\mu\nu}))
&=&\frac{1}{2}\sqrt{h}( K h^{\mu \nu} -K^{\mu\nu}) h^{\lambda \tau} \delta g_{\lambda \tau} 
+\sqrt{h} ( h^{\mu \nu} \delta K + K \delta h^{\mu\nu} -\delta K^{\mu\nu})  \q .\q\q
\ea
and thus need to find $\delta h^{\mu\nu}$ and $\delta K^{\mu\nu}$. We found $\delta K$ in (\ref{C17a}).

 The variation of the inverse boundary metric gives 
\ba
\delta h^{\mu \nu} &=& \delta( g^{\mu \rho} g^{\nu \sigma} (g_{\rho \sigma}-n_\rho n_\sigma)) \nn\\
&=& -h^{\mu \tau} h^{\nu \lambda} \delta g_{\lambda \tau}
\ea
where to go from the first to the second line one uses  $\delta n_\mu = \frac{1}{2}  n_\mu  n^\rho n^\nu \delta g_{\rho \nu} $ and $\delta g^{\mu \rho}=-g^{\mu \lambda} g^{\rho \tau} \delta g_{\lambda \tau}$.

Furthermore we find for the variation of the extrinsic curvature tensor (using again the result for the variation of the normal)
\ba\label{D4}
\delta K^{\mu \nu}
&=&
\delta( h^{\mu\rho} h^{\nu \sigma} \nabla_\rho n_\sigma) \nn\\
&=&
-h^{\mu \lambda} h^{\rho \tau} h^{\nu \sigma} (\nabla_\rho n_\sigma) \delta g_{\lambda \tau} 
-h^{\mu \rho} h^{\nu\lambda} h^{\tau \sigma} (\nabla_\rho n_\sigma) \delta g_{\lambda \tau} 
+ h^{\mu \rho} h^{\nu\sigma}  \nabla_\rho  ( \frac{1}{2} n_\sigma n^\lambda n^\tau \delta g_{\lambda \tau}) \nn\\
&&
-  h^{\mu \rho} h^{\nu\sigma}  \frac{1}{2} n^\kappa ( \nabla_\rho \delta g_{\sigma \kappa} + \nabla_\sigma \delta g_{\sigma \kappa} - \nabla_{\kappa} \delta g_{\rho \sigma})  \q .
\ea
Using that $h^{\nu\sigma}  n_\sigma =0$ for the last term in the second line of (\ref{D4}) we can rewrite this variation into
\begin{equation}
\delta K^{\mu \nu} =
-K^{\tau \nu} h^{\mu \lambda} \delta g_{\lambda \tau} - K^{\mu \tau} h^{\nu \lambda} \delta g_{\lambda \tau}
+ \frac{1}{2} K^{\mu\nu} n^\lambda n^\tau \delta g_{\lambda \tau}  
- \frac{1}{2} h^{ \mu\rho} h^{\nu \sigma} n^\kappa  ( \nabla_\rho \delta g_{\sigma \kappa} + \nabla_\sigma \delta g_{\sigma \kappa} - \nabla_{\kappa} \delta g_{\rho \sigma})  \q .
\end{equation}
We thus obtain
\ba
&&\delta( \sqrt{h}( K h^{\mu \nu} -K^{\mu\nu}))\nn\\
&=&
\sqrt{h}  \left(
\frac{1}{2}K h^{\mu \nu} h^{\lambda \tau} - K h^{\mu \tau} h^{\nu \lambda}  -\frac{1}{2} K^{\mu\nu} g^{\lambda \tau}    -\frac{1}{2} h^{\mu\nu} K^{\lambda \tau}  
+ h^{\mu \lambda} K^{\nu \tau} + h^{\nu \lambda} K^{\mu \tau} \right)\delta g_{\lambda \tau} \nn\\
&&
+\frac{1}{2}\sqrt{h}
\left(  
-h^{\mu \nu}  g^{ \lambda \tau} n^\kappa (\nabla_\lambda \delta g_{\tau \kappa}- \nabla_\kappa \delta g_{ \lambda \tau} )
+ h^{\mu\tau} h^{\nu \lambda} n^\kappa (\nabla_\tau \delta g_{\lambda \kappa} +\nabla_\lambda \delta g_{\tau \kappa}- \nabla_\kappa \delta g_{\lambda \tau} \right)\nn\\
&&-\frac{1}{2}\sqrt{h} h^{\mu\nu} D^\rho w_\rho \q .
\ea
The last term can be rewritten as
\ba
D^\mu w_\mu &=& h^{\mu \nu} \nabla_\mu (h^\rho_\nu n^\lambda \delta g_{\rho \lambda}) \nn\\
&=& h^{ \lambda \tau} n^\kappa  \nabla_\tau \delta g_{\lambda \kappa} - K n^\lambda n^\tau \delta g_{\lambda \tau} + K^{\lambda \tau} \delta g_{\lambda \tau}
\ea
so that we achieved to show
\ba
&&\delta( \sqrt{h}( K h^{\mu \nu} -K^{\mu\nu}))\nn\\
&=&
\sqrt{h}  \left(
\frac{1}{2}(K h^{\mu \nu} -K^{\mu\nu})g^{\lambda \tau} - K h^{\mu \tau} h^{\nu \lambda}    - h^{\mu\nu} K^{\lambda \tau}  
+ h^{\mu \lambda} K^{\nu \tau} + h^{\nu \lambda} K^{\mu \tau} \right)\delta g_{\lambda \tau} \nn\\
&&
+\frac{1}{2}\sqrt{h}
\left(  
(  h^{\mu\tau} h^{\nu \lambda}-h^{\mu \nu}  h^{ \lambda \tau} )n^\kappa (\nabla_\tau \delta g_{\lambda \kappa} +\nabla_\lambda \delta g_{\tau \kappa}- \nabla_\kappa \delta g_{ \lambda \tau} ) \right) 
\ea
where we used $n^\lambda n^\tau n^\kappa (\nabla_\lambda \delta g_{\tau \kappa}- \nabla_\kappa \delta g_{ \lambda \tau} )=0$.
\end{appendix}


\begin{thebibliography}{99}\small
\parskip -1pt

\bibitem{maldacena}
  J.~M.~Maldacena,
  ``The Large N limit of superconformal field theories and supergravity,''
  Int.\ J.\ Theor.\ Phys.\  {\bf 38} (1999) 1113
   [Adv.\ Theor.\ Math.\ Phys.\  {\bf 2} (1998) 231]
  [hep-th/9711200].

\bibitem{witten}
  E.~Witten,
  ``Anti-de Sitter space and holography,''
  Adv.\ Theor.\ Math.\ Phys.\  {\bf 2} (1998) 253
  [hep-th/9802150].
  
  \bibitem{book}
  M. Ammon and J. Erdmenger, ``Gauge/Gravity Duality: Foundations and Applications" (Cambridge University Press, Cambridge 2015)


\bibitem{lqg}
A. Ashtekar, J. Lewandowski, ``Background Independent Quantum Gravity:
A Status Report'', Class. Quant. Grav. {\bf 21} (2004) R53;
[gr-qc/0404018].
  C. Rovelli, ``Quantum Gravity'' (Cambridge University Press,
Cambridge 2004).
 T.~Thiemann,
  ``Modern canonical quantum general relativity,''
  (Cambridge University Press, Cambridge  2007)
  [gr-qc/0110034].


\bibitem{spinfoams}
 A.~Perez,
  ``The Spin Foam Approach to Quantum Gravity,''
  Living Rev.\ Rel.\  {\bf 16} (2013) 3
  [arXiv:1205.2019 [gr-qc]].


\bibitem{freidelAdS} 
 L.~Freidel,
  ``Reconstructing AdS/CFT,''
  arXiv:0804.0632 [hep-th].

\bibitem{KIRILL}
 L.~Freidel and K.~Krasnov,
  ``2-D conformal field theories and holography,''
  J.\ Math.\ Phys.\  {\bf 45} (2004) 2378
  [hep-th/0205091].

\bibitem{BonzomLivineIsing}
V.~Bonzom, F.~Costantino and E.~R.~Livine,
  ``Duality between Spin networks and the 2D Ising model,''
  arXiv:1504.02822 [math-ph].

\bibitem{Norbert}
 N.~Bodendorfer,
  ``A note on quantum supergravity and AdS/CFT,''
  arXiv:1509.02036 [hep-th].
  
  
  \bibitem{girelli} 
  M.~Dupuis and F.~Girelli,
  ``Quantum hyperbolic geometry in loop quantum gravity with cosmological constant,''
  Phys.\ Rev.\ D {\bf 87} (2013) 121502
  [arXiv:1307.5461 [gr-qc]].
 V. Bonzom, M. Dupuis, F. Girelli and E. R. Livine,
``Deformed phase space for 3d loop gravity and hyperbolic discrete geometries'',
arXiv:1402.2323 [gr-qc].

  
  \bibitem{aldo} H. M. Haggard, M. Han, W. Kami\'nski and A. Riello,
``SL(2,C) Chern--Simons Theory, a non-planar graph operator, and 4D loop quantum gravity with a cosmological constant: semi-classical geometry'',
(2014), arXiv:1412.7546 [hep-th].
 H.~M.~Haggard, M.~Han and A.~Riello,
  ``Encoding Curved Tetrahedra in Face Holonomies: a Phase Space of Shapes from Group-Valued Moment Maps,''
  arXiv:1506.03053 [math-ph].
  H.~M.~Haggard, M.~Han, W.~Kami\'nski and A.~Riello,
  ``Four-dimensional Quantum Gravity with a Cosmological Constant from Three-dimensional Holomorphic Blocks,''
  arXiv:1509.00458 [hep-th].
  
\bibitem{BarnichFlat} 
 G.~Arcioni and C.~Dappiaggi,
  ``Exploring the holographic principle in asymptotically flat space-times via the BMS group,''
  Nucl.\ Phys.\ B {\bf 674} (2003) 553
  [hep-th/0306142].
G.~Barnich and G.~Compere,
  ``Classical central extension for asymptotic symmetries at null infinity in three spacetime dimensions,''
  Class.\ Quant.\ Grav.\  {\bf 24} (2007) F15
  [gr-qc/0610130].
   G.~Barnich and C.~Troessaert,
  ``Aspects of the BMS/CFT correspondence,''
  JHEP {\bf 1005} (2010) 062
  [arXiv:1001.1541 [hep-th]].
   G.~Barnich, A.~Gomberoff and H.~A.~Gonzalez,
  ``The Flat limit of three dimensional asymptotically anti-de Sitter spacetimes,''
  Phys.\ Rev.\ D {\bf 86} (2012) 024020
  [arXiv:1204.3288 [gr-qc]].
  
\bibitem{Barnich1} 
G.~Barnich, H.~A.~Gonzalez, A.~Maloney and B.~Oblak,
  ``One-loop partition function of three-dimensional flat gravity,''
  JHEP {\bf 1504} (2015) 178
  [arXiv:1502.06185 [hep-th]].

\bibitem{BMS} 
H.~Bondi, M.~G.~J.~van der Burg and A.~W.~K.~Metzner,
  ``Gravitational waves in general relativity. 7. Waves from axisymmetric isolated systems,''
  Proc.\ Roy.\ Soc.\ Lond.\ A {\bf 269} (1962) 21.
   R.~K.~Sachs,
  ``Gravitational waves in general relativity. 8. Waves in asymptotically flat space-times,''
  Proc.\ Roy.\ Soc.\ Lond.\ A {\bf 270} (1962) 103.
  R.~Sachs,
  ``Asymptotic symmetries in gravitational theory,''
  Phys.\ Rev.\  {\bf 128} (1962) 2851.
  A.~Ashtekar, J.~Bicak and B.~G.~Schmidt,
  ``Asymptotic structure of symmetry reduced general relativity,''
  Phys.\ Rev.\ D {\bf 55} (1997) 669
  [gr-qc/9608042].
 
\bibitem{oeckl}
R.~Oeckl,
``A 'General boundary' formulation for quantum mechanics and quantum gravity,''
Phys.\ Lett.\ B {\bf 575} (2003) 318
[hep-th/0306025].
  
  \bibitem{RovelliHJ}
    C.~Rovelli,
  ``On the structure of a background independent quantum theory: Hamilton function, transition amplitudes, classical limit and continuous limit,''
  arXiv:1108.0832 [gr-qc].
  
  \bibitem{bd12}
 B.~Dittrich,
  ``From the discrete to the continuous: Towards a cylindrically consistent dynamics,''
  New J.\ Phys.\  {\bf 14} (2012) 123004
  [arXiv:1205.6127 [gr-qc]].
  
  \bibitem{bd14} B. Dittrich,
``The continuum limit of loop quantum gravity - a framework for solving the theory'',
in A. Ashtekar and J. Pullin, ed., to be published in the World Scientific series ``100 Years of General Relativity'',
(2014), arXiv:1409.1450 [gr-qc].
  
  \bibitem{PerezRovelli}
  A.~Perez and C.~Rovelli,
  ``Observables in quantum gravity,''
  In 'Quanta of Maths', E.Blanchard, D.Ellwood, M.Khalkhali, M.Marcolli, H.Moscovici, S.Popa eds., 501-518, American Mathematical Society (2011)
  [gr-qc/0104034].
  

 \bibitem{skenderis}
 K.~Skenderis and S.~N.~Solodukhin,
  ``Quantum effective action from the AdS / CFT correspondence,''
  Phys.\ Lett.\ B {\bf 472} (2000) 316
  [hep-th/9910023].
  S.~de Haro, S.~N.~Solodukhin and K.~Skenderis,
  ``Holographic reconstruction of space-time and renormalization in the AdS / CFT correspondence,''
  Commun.\ Math.\ Phys.\  {\bf 217} (2001) 595
  [hep-th/0002230].
 
 \bibitem{carlip}
S.~Carlip,
  ``Dynamics of asymptotic diffeomorphisms in (2+1)-dimensional gravity,''
  Class.\ Quant.\ Grav.\  {\bf 22} (2005) 3055
  [gr-qc/0501033].
    S.~Carlip,
  ``Conformal field theory, (2+1)-dimensional gravity, and the BTZ black hole,''
  Class.\ Quant.\ Grav.\  {\bf 22} (2005) R85
  [gr-qc/0503022].
  

 \bibitem{regge}
  T.~Regge,
  ``General relativity without coordinates,''
  Nuovo Cim.\  {\bf 19} (1961) 558.
 
 \bibitem{qregge}
 T.~Regge and R.~M.~Williams,
  ``Discrete structures in gravity,''
  J.\ Math.\ Phys.\  {\bf 41} (2000) 3964
  [arXiv:gr-qc/0012035].
H.~W.~Hamber,
  ``Quantum gravitation: The Feynman path integral approach,''
  (Berlin, Germany: Springer (2009) )
  
 


\bibitem{Meas}
B.~Dittrich and S.~Steinhaus,
  ``Path integral measure and triangulation independence in discrete gravity,''
  Phys.\ Rev.\ D {\bf 85} (2012) 044032
  [arXiv:1110.6866 [gr-qc]].
   B.~Dittrich, W.~Kaminski and S.~Steinhaus,
  ``Discretization independence implies non-locality in 4D discrete quantum gravity,''
  arXiv:1404.5288 [gr-qc].

\bibitem{PR}
G. Ponzano and T. Regge, ``Semiclassical limit of racah coefficients,'' in {\it Spectroscopy and
group theoretical methods in physics} (F. Block, ed.), North Holland, 1968.
  L.~Freidel and D.~Louapre,
  ``Ponzano-Regge model revisited. I: Gauge fixing, observables and
  interacting spinning particles,''
  Class.\ Quant.\ Grav.\  {\bf 21} (2004) 5685
  [arXiv:hep-th/0401076].
  

\bibitem{TorsionPR}
  L.~Freidel and D.~Louapre,
  ``Ponzano-Regge model revisited II: Equivalence with Chern-Simons,''
  gr-qc/0410141.
  J.~W.~Barrett and I.~Naish-Guzman,
  ``The Ponzano-Regge model,''
  Class.\ Quant.\ Grav.\  {\bf 26} (2009) 155014
  [arXiv:0803.3319 [gr-qc]].
  V.~Bonzom and M.~Smerlak,
  ``Bubble divergences from twisted cohomology,''
  Commun.\ Math.\ Phys.\  {\bf 312}, 399 (2012)
  [arXiv:1008.1476 [math-ph]].
  V.~Bonzom and M.~Smerlak,
  ``Gauge symmetries in spinfoam gravity: the case for 'cellular quantization',''
  Phys.\ Rev.\ Lett.\  {\bf 108}, 241303 (2012)
  [arXiv:1201.4996 [gr-qc]].

\bibitem{FG}
C. Fefferman and C. R. Graham, ``Conformal Invariants,"  in Elie Cartan et les Math\'ematiques d'aujourd'hui (Asterisque, 1985) 95

\bibitem{MERA}
 ``Entanglement Renormalization,''
  Phys.\ Rev.\ Lett.\  {\bf 99} (2007) 22,  220405
  [cond-mat/0512165].
  G.~Evenbly and G.~Vidal,
  ``Entanglement renormalization in two spatial dimensions,''
  Phys.\ Rev.\ Lett.\  {\bf 102} (2009) 180406
  [arXiv:0811.0879 [cond-mat.str-el]].

\bibitem{AdSMERA}
B.~Swingle,
  ``Constructing holographic spacetimes using entanglement renormalization,''
  arXiv:1209.3304 [hep-th].

\bibitem{dectnw}
B. Dittrich, S. Mizera and S. Steinhaus,
``Decorated tensor network renormalization for lattice gauge theories and spin foam models'',
(2014), arXiv:1409.2407 [gr-qc].  

\bibitem{improved}
B. Bahr and B. Dittrich, 
``Improved and Perfect Actions in Discrete Gravity'',
Phys. Rev. D \textbf{80} (2009) 124030, [arXiv:0907.4323 [gr-qc]].

\bibitem{perfect}
 B. Bahr, B. Dittrich and S. Steinhaus,
``Perfect discretization of reparametrization invariant path integrals'',
Phys. Rev. D \textbf{83} (2011) 105026, [arXiv:1101.4775 [gr-qc]].


\bibitem{ThermalSpinning}
G.~Barnich,
  ``Entropy of three-dimensional asymptotically flat cosmological solutions,''
  JHEP {\bf 1210} (2012) 095
  [arXiv:1208.4371 [hep-th]].

\bibitem{GHY} 
G.~W.~Gibbons and S.~W.~Hawking,
  ``Action Integrals and Partition Functions in Quantum Gravity,''
  Phys.\ Rev.\ D {\bf 15} (1977) 2752.
  J.~W.~York, Jr.,
  ``Role of conformal three geometry in the dynamics of gravitation,''
  Phys.\ Rev.\ Lett.\  {\bf 28} (1972) 1082.


\bibitem{detournay}
 A.~Bagchi, S.~Detournay, D.~Grumiller and J.~Simon,
  ``Cosmic Evolution from Phase Transition of Three-Dimensional Flat Space,''
  Phys.\ Rev.\ Lett.\  {\bf 111} (2013) 18,  181301
  [arXiv:1305.2919 [hep-th]].
S.~Detournay, D.~Grumiller, F.~Sch\"oller and J.~Simon,
  ``Variational principle and one-point functions in three-dimensional flat space Einstein gravity,''
  Phys.\ Rev.\ D {\bf 89} (2014) 8,  084061
  [arXiv:1402.3687 [hep-th]].

\bibitem{giombi}
 S.~Giombi, A.~Maloney and X.~Yin,
  ``One-loop Partition Functions of 3D Gravity,''
  JHEP {\bf 0808} (2008) 007
  [arXiv:0804.1773 [hep-th]].

\bibitem{oblak}
B.~Oblak,
  ``Characters of the BMS Group in Three Dimensions,''
  Commun.\ Math.\ Phys.\  {\bf 340} (2015) 1,  413
  [arXiv:1502.03108 [hep-th]].



\bibitem{sorkinBdry}
J.~B.~Hartle and R.~Sorkin,
  ``Boundary Terms in the Action for the Regge Calculus,''
  Gen.\ Rel.\ Grav.\  {\bf 13} (1981) 541.

  \bibitem{rocek} 
M.~Rocek and R.~M.~Williams,
  ``Quantum Regge Calculus,''
  Phys.\ Lett.\  B {\bf 104} (1981) 31.
 M.~Rocek and R.~M.~Williams,
  ``The Quantization Of Regge Calculus,''
  Z.\ Phys.\  C {\bf 21} (1984) 371.
  
  \bibitem{dittrich08}
  B.~Dittrich,
  ``Diffeomorphism symmetry in quantum gravity models,''
  Adv.\ Sci.\ Lett.\  {\bf 2} 151
  [arXiv:0810.3594 [gr-qc]].
   B.~Bahr and B.~Dittrich,
  ``(Broken) Gauge Symmetries and Constraints in Regge Calculus,''
  Class.\ Quant.\ Grav.\  {\bf 26} (2009) 225011
  [arXiv:0905.1670 [gr-qc]].
  B.~Bahr and B.~Dittrich,
  ``Breaking and restoring of diffeomorphism symmetry in discrete gravity,''
  AIP Conf.\ Proc.\  {\bf 1196} (2009) 10
  [arXiv:0909.5688 [gr-qc]].
    
 \bibitem{bd12review}
B.~Dittrich,
  ``How to construct diffeomorphism symmetry on the lattice,''
  PoS QGQGS {\bf 2011} (2011) 012
  [arXiv:1201.3840 [gr-qc]].
 
 \bibitem{he}
B.~Bahr, B.~Dittrich, S.~He,
  ``Coarse graining free theories with gauge symmetries: the linearized case,''
  New J.\ Phys.\  {\bf 13 } (2011)  045009,
[arXiv:1011.3667 [gr-qc]].

  
  
\bibitem{hoehn2}
B.~Dittrich and P.~A.~H\"ohn,
``From covariant to canonical formulations of discrete gravity,''
Class.\ Quant.\ Grav.\  {\bf 27} (2010) 155001
[arXiv:0912.1817 [gr-qc]].
%
B.~Dittrich and P.~A.~H\"ohn,
``Canonical simplicial gravity,''
Class.\ Quant.\ Grav.\  {\bf 29} (2012) 115009
[arXiv:1108.1974 [gr-qc]].
%
%
B.~Dittrich and P.~A.~H\"ohn,
``Constraint analysis for variational discrete systems,''
J.\  Math.\  Phys.\  {\bf 54}   (2013) 093505
[arXiv:1303.4294 [math-ph]].

\bibitem{antispacetime}
  M.~Christodoulou, M.~Langvik, A.~Riello, C.~Roken and C.~Rovelli,
  ``Divergences and Orientation in Spinfoams,''
  Class. Quantum Grav. 30 055009 2013
  [arXiv:1207.5156 [gr-qc]].


\bibitem{hamber-williams} 
H.~W.~Hamber and R.~M.~Williams,
  ``Gauge invariance in simplicial gravity,''
  Nucl.\ Phys.\  B {\bf 487} (1997) 345
  [arXiv:hep-th/9607153].


 
 \bibitem{louapre} 
L.~Freidel and D.~Louapre,
  ``Diffeomorphisms and spin foam models,''
  Nucl.\ Phys.\  B {\bf 662} (2003) 279
  [arXiv:gr-qc/0212001].
  

\bibitem{dfs}
 B.~Dittrich, L.~Freidel and S.~Speziale,
  ``Linearized dynamics from the 4-simplex Regge action,''
  Phys.\ Rev.\ D {\bf 76} (2007) 104020
  [arXiv:0707.4513 [gr-qc]].


\bibitem{Reggemeasure}
F.~Lund and T.~Regge, 1974, unpublished. 
 P.~Menotti and P.~P.~Peirano,
  ``Diffeomorphism invariant measure for finite dimensional geometries,''
  Nucl.\ Phys.\  B {\bf 488} (1997) 719
[arXiv:hep-th/9607071].
J.~Ambjorn, J.~L.~Nielsen, J.~Rolf and G.~K.~Savvidy,
  ``Spikes in quantum Regge calculus,''
  Class.\ Quant.\ Grav.\  {\bf 14} (1997) 3225
[arXiv:gr-qc/9704079].
 H.~W.~Hamber, R.~M.~Williams,
  ``On the measure in simplicial gravity,''
  Phys.\ Rev.\  {\bf D59 } (1999)  064014.
[hep-th/9708019].



  \bibitem{asympPR}
J.~W.~Barrett and T.~J.~Foxon,
``Semiclassical limits of simplicial quantum gravity,''
Class.\ Quant.\ Grav.\  {\bf 11} (1994) 543
[gr-qc/9310016].
  J.~Roberts, ``Classical 6j-symbols and the tetrahedron,'' Geom.\ Topol.\. {\bf  3} (1999) 21.
1999.

\bibitem{GibbonsHawking}
G.~W.~Gibbons, S.~W.~Hawking, M.~J.~Perry,
  ``Path Integrals and the Indefiniteness of the Gravitational Action,''
  Nucl.\ Phys.\  {\bf B138 } (1978)  141.


\bibitem{aristide3D}
 A.~Baratin and L.~Freidel,
  ``Hidden Quantum Gravity in 3-D Feynman diagrams,''
  Class.\ Quant.\ Grav.\  {\bf 24} (2007) 1993,
[arXiv:gr-qc/0604016].

\bibitem{sorkin}
  R.~Sorkin,
  ``The Electromagnetic field on a simplicial net,''
  J.\ Math.\ Phys.\  {\bf 16}, 2432 (1975)
  [Erratum-ibid.\  {\bf 19}, 1800 (1978)].

\bibitem{barnichLiou}
G.~Barnich, A.~Gomberoff and H.~A.~Gonzalez,
  ``Three-dimensional Bondi-Metzner-Sachs invariant two-dimensional field theories as the flat limit of Liouville theory,''
  Phys.\ Rev.\ D {\bf 87} (2013) 12,  124032
  [arXiv:1210.0731 [hep-th]].

\bibitem{bdsteinhaus13}
 B.~Dittrich and S.~Steinhaus,
  ``Time evolution as refining, coarse graining and entangling,''
  New J.\ Phys.\  {\bf 16} (2014) 123041
  [arXiv:1311.7565 [gr-qc]].

\bibitem{AltMERA}
B.~Czech, L.~Lamprou, S.~McCandlish and J.~Sully,
  ``Integral Geometry and Holography,''
  JHEP {\bf 1510} (2015) 175
  [arXiv:1505.05515 [hep-th]].

\bibitem{HolR}
M.~Bianchi, D.~Z.~Freedman and K.~Skenderis,
  ``Holographic renormalization,''
  Nucl.\ Phys.\ B {\bf 631} (2002) 159
  [hep-th/0112119].
 K.~Skenderis,
  ``Lecture notes on holographic renormalization,''
  Class.\ Quant.\ Grav.\  {\bf 19} (2002) 5849
  [hep-th/0209067].

\bibitem{barret}
 J.~W.~Barrett,
  ``First order Regge calculus,''
  Class.\ Quant.\ Grav.\  {\bf 11} (1994) 2723
  [hep-th/9404124].

\bibitem{newangle}
B. Bahr and B. Dittrich,
``Regge calculus from a new angle'',
New J. Phys. \textbf{12} 033010 (2010), [arXiv:0907.4325 [gr-qc]].


\bibitem{areaangle}
B.~Dittrich and S.~Speziale,
  ``Area-angle variables for general relativity,''
  New J.\ Phys.\  {\bf 10} (2008) 083006
  [arXiv:0802.0864 [gr-qc]].

\bibitem{AEtoappear}
E.~Livine, A.~Riello, w.i.p.

\bibitem{eckert}
 B.~Dittrich, F.~C.~Eckert and M.~Martin-Benito,
  ``Coarse graining methods for spin net and spin foam models,''
  New J.\ Phys.\  {\bf 14} (2012) 035008
  [arXiv:1109.4927 [gr-qc]].
   B.~Dittrich, M.~Martin-Benito and E.~Schnetter,
  ``Coarse graining of spin net models: dynamics of intertwiners,''
  New J.\ Phys.\  {\bf 15} (2013) 103004
  [arXiv:1306.2987 [gr-qc]].
   B.~Dittrich, M.~Martin-Benito and S.~Steinhaus,
  ``Quantum group spin nets: refinement limit and relation to spin foams,''
  Phys.\ Rev.\ D {\bf 90} (2014) 024058
  [arXiv:1312.0905 [gr-qc]].
  B.~Bahr and S.~Steinhaus,
  ``Quantum Cuboids and the EPRL-FK path integral for quantum gravity,''
  arXiv:1508.07961 [gr-qc].

\bibitem{Hellmann}
 R.~J.~Dowdall, H.~Gomes and F.~Hellmann,
  ``Asymptotic analysis of the Ponzano-Regge model for handlebodies,''
  J.\ Phys.\ A  {\bf 43} (2010) 115203,
 [arXiv:0909.2027 [gr-qc]].

\bibitem{BDHnybida}
B.~Dittrich and J.~Hnybida,
  ``Ising Model from Intertwiners,''
  arXiv:1312.5646 [gr-qc].

\bibitem{Witten3D}
  E.~Witten,
  ``(2+1)-Dimensional Gravity as an Exactly Soluble System,''
  Nucl.\ Phys.\  B {\bf 311}, 46 (1988).
  
\bibitem{TorsionBF}
  E.~Witten,
  ``Topology Changing Amplitudes in (2+1)-Dimensional Gravity,''
  Nucl.\ Phys.\  B {\bf 323}, 113 (1989).
  M.~Blau and G.~Thompson,
  ``A new class of topological field theories and the Ray-Singer torsion,''
  Phys.\ Lett.\ B {\bf 228}, 64 (1989).
  S.~Carlip, R.~Cosgrove,
  ``Topology change in (2+1)-dimensional gravity,''
  J.\ Math.\ Phys.\  {\bf 35}, 5477-5493 (1994).
  [gr-qc/9406006].

\bibitem{btz}
M.~Banados, C.~Teitelboim and J.~Zanelli,
  ``The Black hole in three-dimensional space-time,''
  Phys.\ Rev.\ Lett.\  {\bf 69} (1992) 1849
  [hep-th/9204099].

\bibitem{marc}
M.~Geiller and K.~Noui,
  ``BTZ Black Hole Entropy and the Turaev-Viro model,''
  Annales Henri Poincare {\bf 16} (2015) 2,  609
  [arXiv:1312.1696 [gr-qc]].

\bibitem{ABS}
B.~Dittrich, S.~Mizera, A.~Riello, to appear

\bibitem{krasnovPP}
K.~Krasnov,
  ``3-D gravity, point particles and Liouville theory,''
  Class.\ Quant.\ Grav.\  {\bf 18} (2001) 1291
  [hep-th/0008253].


\bibitem{BarrettCrane}
J.~W.~Barrett and L.~Crane,
  ``An Algebraic interpretation of the Wheeler-DeWitt equation,''
  Class.\ Quant.\ Grav.\  {\bf 14} (1997) 2113
  [gr-qc/9609030].

\bibitem{BonzomFreidel}
V.~Bonzom and L.~Freidel,
  ``The Hamiltonian constraint in 3d Riemannian loop quantum gravity,''
  Class.\ Quant.\ Grav.\  {\bf 28} (2011) 195006
  [arXiv:1101.3524 [gr-qc]].

\bibitem{tent}
 J.~W.~Barrett, M.~Galassi, W.~A.~Miller, R.~D.~Sorkin, P.~A.~Tuckey and R.~M.~Williams,
  ``A Paralellizable implicit evolution scheme for Regge calculus,''
  Int.\ J.\ Theor.\ Phys.\  {\bf 36} (1997) 815
  [arXiv:gr-qc/9411008].


\bibitem{Marolf}  T.~Andrade and D.~Marolf,
  ``Asymptotic Symmetries from finite boxes,''
  arXiv:1508.02515 [gr-qc].

\bibitem{bonzomdittrich}
 V.~Bonzom and B.~Dittrich,
  ``Dirac's discrete hypersurface deformation algebras,''
  Class.\ Quant.\ Grav.\  {\bf 30} (2013) 205013
  [arXiv:1304.5983 [gr-qc]].









   
 
 
 
\end{thebibliography}
\end{document}